\documentclass[12pt]{article}
\pdfoutput=1
\usepackage{tikz}
\usetikzlibrary{matrix,arrows,decorations.pathmorphing,decorations.pathreplacing}
\usepackage{jheppub}
\usetikzlibrary{arrows}
\usepackage{amsmath,amssymb,euscript,array,mathrsfs,appendix,ctable,marvosym}
\usepackage{mathtools}
\usepackage{changepage} 
\usepackage{arydshln}
\usepackage{todonotes}
\usepackage{graphicx}
\usepackage{color}

\usepackage[full]{textcomp} 
     \usepackage{fbb} 
          \usepackage[scaled=.95]{cabin}
     \usepackage[varqu,varl]{inconsolata}
     \usepackage[libertine,bigdelims,vvarbb]{newtxmath}
     \usepackage[cal=boondoxo]{mathalfa}
    \usepackage[T1]{fontenc}
\useosf
\newcommand\BLUE[1]{{\color{blue}{#1}}}
\def\UP{z^+}
\def\DOWN{z^-}
\def\SUB#1{\raisebox{-5pt}{$\scriptstyle#1$}}
\def\Ba{{\boldsymbol a}}
\def\Bb{{\boldsymbol b}}
\def\Bc{{\boldsymbol c}}
\def\Bu{{\boldsymbol u}}

\def\FF{{\EuScript F}}
\def\WW{{\EuScript W}}
\def\SS{{\EuScript S}}
\def\EE{{{\EuScript S}^\perp}}
\def\QQ{{\EuScript Q}}

\def\MM{{\EuScript M}}
\def\NN{{\EuScript N}}

\def\BH{{\cal H}}

\newcommand{\Tr}{\operatorname{Tr}}
\newcommand{\IM}{\operatorname{Im}}
\newcommand{\RE}{\operatorname{Re}}

\def\bra#1{\langle #1|}
\def\ket#1{| #1\rangle}

\newcommand{\EQ}[1]{\begin{equation}\begin{split} #1
\end{split}\end{equation}}

\title{Born's rule yields classical states and classical mechanics}
\author{Timothy J. Hollowood}
\affiliation{
Department of Physics, Swansea University,\\ Swansea, SA2 8PP, United Kingdom
}

\emailAdd{t.hollowood@swansea.ac.uk}

\abstract{It is shown that Schr\"odinger's equation and Born's rule are sufficient to ensure that the states of macroscopic collective coordinate subsystems are microscopically localized in phase space and that the localized state follows the classical trajectory with random quantum noise that is indistinguishable from the pseudo-random noise of classical Brownian motion. This happens because in realistic systems the localization rate determined by the coupling to the environment is greater than the Lyapunov exponent that governs chaotic spreading in phase space. For realistic systems, the trajectories of the collective coordinate subsystem are at the same time an ``unravelling'' and a set of ``consistent/decoherent histories''. Different subsystems have their own stochastic dynamics that generally knit together to form a global dynamics, although in certain contrived thought experiments, most notably Wigner's friend, on the contrary, there is observer complementarity.}

\notoc
\begin{document}

\maketitle

\newpage

\section{Introduction}

Quantum Mechanics is often thought to need an ``interpretation'' to order to solve the measurement problem. The latter is really part of a more general problem that we can pose as 2 questions:
\begin{quote}
{\bf Question 1.} How does quantum mechanics ensure that macroscopic systems are in classical states, i.e.~single points in phase space, at least up to the limits imposed by the uncertainty relations?

{\bf Question 2.} Why do the points follow trajectories determined by Newton's laws of motion?
\end{quote}
The answer to the second question follows rather easily, via Ehrenfest's Theorem, if the first question is answered, so the more fundamental problem of the quantum-to-classical transition is to explain the origin of classical states.

Quantum mechanics already has an interpretation in the form of Born's rule which seems to be sufficient for all applications. The present work is concerned with the issue of whether the Born rule, by itself, is sufficient by itself to answer the 2 questions above. If yes, then all the seemingly endless talk of  interpretations of quantum mechanics, as if there were a menu of viable options, would be rendered redundant.\footnote{At the very least, a putative interpretation must answer the questions above, as well as being consistent with locality and causality given that at a fundamental level the world is described by relativistic quantum field theory.}

In standard quantum mechanics there are effective rules that describe measurements. It is here that probabilities as embodied in Born's rule appear in the theory. The textbook version says that if an observable $A$ is measured on a quantum system $\SS$ in the state $\ket{\psi}$, then if we write the state in terms of the eigenvectors of $A$, $A\ket{a}=\xi_a\ket{a}$,
\EQ{
\ket{\psi}=\sum_ac_a\ket{a}\ ,
}
then the outcome $\xi_a$ occurs with probability $|c_a|^2$ and the final state is the eigenstate $\ket{a}$. We call the latter the {\it conditioned state\/} in that it is picked out as the random outcome revealed when the measurement is completed:\footnote{We could equally well refer to the conditioned state as the {\it reduced\/} state or even the {\it collapsed\/} state.} 
\begin{center}
\begin{tikzpicture}[scale=0.7]
\draw[fill=black] (1,0) circle (0.3cm);
\draw[thick] (-1,0)  -- (1,0);
\draw[thick] (1,0)  -- (3,1.5);
\draw[thick] (1,0)  -- (3,0.5);
\draw[thick] (1,0)  -- (3,-0.5);
\draw[thick] (1,0)  -- (3,-1.5);
\node at (-2,0) {$\ket{\psi}$};
\node at (4,0.5) {$\ket{a}$};
\node at (6.5,0.5) {$p_a=|c_a|^2$};
\end{tikzpicture}
\end{center}
It is important that these measurement rules are really just phenomenological rules that apply when a macroscopic system---the measuring device---interacts with the microscopic system $\SS$ and the focus is put on the behaviour of $\SS$. The phenomenological description is all that is needed to use quantum mechanics in practice. 

If one wants to go beyond the effective measurement rules, the measuring device itself should be included in the story and treated according to the rules of quantum mechanics. To this end, let us introduce an idealized measuring device $\MM$ with a pre-measurement state $\ket{\emptyset}\SUB{\MM}$ and post measurement states $\ket{a}\SUB{\MM}$, indicating the outcome $\xi_a$, in the sense that unitary evolution with an eigenstate $\ket{a}$ as the initial state of $\SS$ takes the form
\EQ{
U(t_1,t_0)\ket{a}\SUB{\SS}\otimes\ket{\emptyset}\SUB{\MM}=\ket{a}\SUB{\SS}\otimes\ket{a}\SUB{\MM}\ .
}
Linearity then ensures that
\EQ{
U(t_1,t_0)\Big(\sum_a c_a\ket{a}\SUB{\SS}\Big)\otimes\ket{\emptyset}\SUB{\MM}=\sum_ac_a\ket{a}\SUB{\SS}\otimes\ket{a}\SUB{\MM}\ ,
\label{grr}
}
an entangled state of the microscopic system $\SS$ with the measuring device $\MM$. Note that the effective description is coarse grained in time in the sense that the interaction between $\SS$ and $\MM$ occurs over some brief interval of time in the interval $[t_0,t_1]$.

The question is how the effective measurement rules of $\SS$ arise out of this quantum description of the total system $\SS\cup\MM$? Since the effective description just involves $\SS$ alone, this means that $\SS$ defines a {\it local frame of reference\/} within the total system. The state within this frame, is the 
density operator obtained  by tracing over $\MM$:
\EQ{
\rho\SUB{\SS}=\sum_a|c_a|^2\ket{a}\bra{a}\SUB{\SS}\ .
}
Born's rule motivates the idea that in the frame associated to $\SS$, entanglement corresponds to randomness, so $\rho\SUB{\SS}$ describes an ensemble of the pure states $\ket{a}\SUB{\SS}$ with probability $|c_a|^2$. As long as the $|c_a|^2$ are non-degenerate, this basis is uniquely defined. Equivalently, we can describe the states via the Schmidt decomposition of the total state in \eqref{grr}.\footnote{In the Schmidt decomposition, the 2 sets of states $\{\ket{a}\SUB{\SS}\}$ and $\{\ket{a}\SUB{\SS^\perp}\}$ are both orthonormal.}

\section{Frames and trajectories}

The previous description of measurement is only toy model, but it reveals the key idea that entanglement in a local subsystem frame {\it is\/} randomness. This is the main lesson we carry forward to the case of an arbitrary subsystem and, in particular, a macroscopic subsystem. 

The first task is to identity a suitable local frame to describe a macroscopic system. At some level there is a fundamental Hilbert space $\BH$. However, in order to describe the behaviour of the system at macroscopic scales, we need to focus on the low-energy collective excitations of the system. These can be expected to define a classical configuration space and the quantum system we are after is the quantization of this classical system. We will call this collective coordinate system $\SS$ since, although it is now macroscopic, it will play the same r\^ole as the microscopic system in the last section. We will denote the Hilbert space of the collective coordinates as $\BH_\SS$. This Hilbert space, tensored with a suitable complement $\BH_{\EE}$---the environment---is a subspace of the total $\BH_{\SS\cup\EE}\equiv\BH_\SS\otimes\BH_{\EE}\subset\BH$. There is some arbitrariness here, as expected in a phenomenological theory, but we choose the environmental Hilbert space to be big enough so that quantum state of $\SS$ is {\it purified\/} on $\BH_{\SS\cup\EE}$ and time evolution on it is approximately unitary, 
\EQ{
\ket{\Psi(t)}=U(t,t_0)\ket{\Psi(t_0)}\ ,\qquad\text{for}\qquad \ket{\Psi(t)}\in\BH_{\SS\cup\EE}\ . 
}
The key point here, is that the evolution on $\BH_\SS$ alone is not unitary due to the fact that the collective coordinate system $\SS$ is interacting with its environment $\EE$, but taken together the evolution on $\SS\cup\EE$ is approximately unitary.

The evolution of the state $\ket{\Psi(t)}$ is unitary but what is relevant to the subsystem frame $\SS$ is the evolution of the {\it conditioned state\/} $\ket{\psi}\in\BH_\SS$ defined as follows. A realistic environment $\EE$ of a macroscopic system can be viewed as interacting with $\SS$ via a series of discrete scattering events that occur over some microscopic time scale $\Delta t$ and repeat over a larger, but still microscopic, time scale $\delta t>\Delta t$. As an example, we will consider the case when $\SS$ describes the position of a macroscopic particle and we will ignore the internal collective coordinates of the particle. A realistic environment could be external, consisting of gas molecules, or photons, which scatter off the particle in individual events. Macroscopic objects also have an internal environment of phonons that interact with their collective coordinates when the particle is subject to external forces. Essentially, each scattering event can be viewed as an irreducible micro-measurement made by the environment on the particle. So each scattering event builds up entanglement as in \eqref{grr}. In the frame of  $\SS$, this becomes the randomness of the conditioned state according to Born's rule. Each scattering event has a very weak effect on the state, in the sense that, in the sum \eqref{grr}, only one of the terms has an appreciable probability $\approx1$.\footnote{The coarse graining here is a key feature of the stochastic dynamics we are defining. If we tried to define the dynamics in a continuum limit $\delta t\to0$ then we run up against the quantum zeno effect and the process would become trivial in the sense that as $\delta t\to0$, one of the conditional probabilities goes like $1-{\cal O}(\delta t^2)$ while all the others are ${\cal O}(\delta t^2)$. So the time scale of each scattering event $\Delta t$ defines a lower bound on the temporal discreteness of the stochastic process. On the other hand, we will find that since each scattering event has a very weak effect, the stochastic process is insensitive to taking the scale $\delta t$ much larger so that many scattering events occur within each interval $\delta t$.}

At the coarse-grained level, the evolution of the conditioned state can therefore be described as a discrete trajectory $\ket{\psi(t_n)}=\ket{\Ba^{(n)}}\SUB{\SS}$:
\EQ{
\ket{\Ba^{(0)}}\SUB{\SS}\longrightarrow \ket{\Ba^{(1)}}\SUB{\SS}\longrightarrow \cdots\longrightarrow \ket{\Ba^{(n-1)}}\SUB{\SS}\longrightarrow \ket{\Ba^{(n)}}\SUB{\SS}\longrightarrow\cdots\ ,
}
specified by the vector label $\Ba^{(n)}=(a_1,a_2,\ldots,a_n)$ at times
$t=n\delta t$, where each label $a_j$ ranges generically over the dimension of $\BH_\SS$, but which in realistic cases will range over a much smaller set. 

The trajectory of the conditioned state is schematically of the form:
\begin{center}
\begin{tikzpicture}[scale=0.6]
\draw[fill=black] (0,0) circle (0.3cm);
\draw[thick] (-3,0)  -- (0,0);
\draw[thick] (0,0)  -- (3,1.5);
\draw[thick] (0,0)  -- (3,0.5);
\draw[thick] (0,0)  -- (3,-0.5);
\draw[thick] (0,0)  -- (3,-1.5);
\node at (2.3,0.7) {\small $a_1$};
\begin{scope}[xshift=3cm,yshift=0.5cm]
\draw[fill=black] (0,0) circle (0.3cm);
\draw[thick] (0,0)  -- (3,1.5);
\draw[thick] (0,0)  -- (3,0.5);
\draw[thick] (0,0)  -- (3,-0.5);
\draw[thick] (0,0)  -- (3,-1.5);
\node at (2.3,-0.1) {\small $a_2$};
\begin{scope}[xshift=3cm,yshift=-0.5cm]
\draw[fill=black] (0,0) circle (0.3cm);
\draw[thick] (0,0)  -- (3,1.5);
\draw[thick] (0,0)  -- (3,0.5);
\draw[thick] (0,0)  -- (3,-0.5);
\draw[thick] (0,0)  -- (3,-1.5);
\node at (2.3,1.7) {\small $a_3$};
\begin{scope}[xshift=3cm,yshift=1.5cm]
\draw[fill=black] (0,0) circle (0.3cm);
\draw[thick] (0,0)  -- (3,1.5);
\draw[thick] (0,0)  -- (3,0.5);
\draw[thick] (0,0)  -- (3,-0.5);
\draw[thick] (0,0)  -- (3,-1.5);
\node at (2.3,0.7) {\small $a_4$};
\begin{scope}[xshift=6cm,yshift=0.5cm]
\draw[dotted,thick] (0,0)  -- (-3,0);
\draw[fill=black] (0,0) circle (0.3cm);
\draw[thick] (0,0)  -- (3,1.5);
\draw[thick] (0,0)  -- (3,0.5);
\draw[thick] (0,0)  -- (3,-0.5);
\draw[thick] (0,0)  -- (3,-1.5);
\node at (2.3,-0.1) {\small $a_n$};
\node[right] at (-4,-3.5) {$\boxed{\Ba^{(n)}=(a_1,a_2,\ldots,a_n)}$};
\end{scope}
\end{scope}
\end{scope}
\end{scope}
\end{tikzpicture}
\end{center}

At any given time after which there have been $n-1$ scattering events leading to a conditioned state $\ket{\Ba^{(n-1)}}\SUB{\SS}$, an additional scattering event corresponds to a lengthening of the label $\Ba^{(n-1)}\to\Ba^{(n)}$ described by the decomposition of the state as in \eqref{grr}:
\EQ{
\boxed{U(t_n,t_{n-1})\ket{\Ba^{(n-1)}}\SUB{\SS}\otimes\ket{\Ba^{(n-1)}}\SUB{\EE}=\sum_{a_{n}}c_{\Ba^{(n)}}\ket{\Ba^{(n)}}\SUB{\SS}\otimes\ket{\Ba^{(n)}}\SUB{\EE}\ ,}
\label{kek2}
}
with a sum over the last component of $\Ba^{(n)}=(a_1,\ldots,a_n)$.
Born's rule dictates the orthogonality conditions in the last index $a_n$:
\EQ{
\bra{\Ba^{(n)}}\Ba^{\prime(n)}\rangle\SUB{\SS}=\bra{\Ba^{(n)}}\Ba^{\prime(n)}\rangle\SUB{\EE}=\delta_{a_na_n'}\ ,
\label{run}
}
where $\Ba^{(n)}=(a_1,\ldots,a_{n-1},a_n)$ and $\Ba^{\prime(n)}=(a_1,\ldots,a_{n-1},a'_n)$ so that \eqref{kek2} is a Schmidt decomposition. Born's rule implies that the conditional probability for $\ket{\Psi_{\Ba^{(n-1)}}}\longrightarrow\ket{\Psi_{\Ba^{(n)}}}$, 
where 
\EQ{
\ket{\Psi_{\Ba^{(n)}}}\equiv\ket{\Ba^{(n)}}\SUB{\SS}\otimes\ket{\Ba^{(n)}}\SUB{\EE}\ ,
} 
is equal to  $|c_{\Ba^{(n)}}|^2$. The probabilities for a trajectory is then obtained by multiplying these together along the trajectory
\EQ{
p_{\Ba^{(n)}}=\big|c_{\Ba^{(n)}}c_{\Ba^{(n-1)}}\cdots c_{\Ba^{(1)}}\big|^2\ ,\qquad \sum_{\Ba^{(n)}}p_{\Ba^{(n)}}=1\ .
\label{tpr}
}
The total state is written as 
\EQ{
\ket{\Psi(t_n)}=\sum_{\Ba^{(n)}}c_{\Ba^{(n)}}c_{\Ba^{(n-1)}}\cdots c_{\Ba^{(1)}}\ket{\Ba^{(n)}}\SUB{\SS}\otimes\ket{\Ba^{(n)}}\SUB{\EE}\ .
\label{lol}
}

\vspace{0.2cm}\noindent
{\bf Decoherence:\/} for a general frame, the states $\ket{\Psi_{\Ba^{(n)}}}$ are not orthogonal.
However, there are additional conditions that apply for realistic macroscopic subsystems. The point is that the environment is a very large quantum system that interacts locally with $\SS$ and rapidly disperses the entanglement resulting from a scattering event away from $\SS$. Another way to say this is that the environment effectively always presents a fresh portion of itself to interact with $\SS$ with no previous correlation. This is the Born-Markov property which leads to the {\it decoherence condition\/} that implies that the states $\ket{\Ba^{(n)}}\SUB{\EE}$ of the environment are orthogonal on {\it all\/} the elements of the vector $\Ba^{(n)}$:
\EQ{
\text{\small\bf Decoherence condition:}\qquad\boxed{\bra{\Ba^{(n)}}\Bb^{(n)}\rangle\SUB{\EE}=\delta_{\Ba^{(n)}\Bb^{(n)}}\equiv\delta_{a_1b_1}\delta_{a_2b_2}\cdots\delta_{a_nb_n} \ .}\qquad\qquad\phantom{.}
\label{kek}
}
Macroscopic frames of low-energy collective coordinate subsystems can be expected to define such decoherent frames, at least to a very high precision. Ultimately, the only frames that have any relevance are those macroscopic ones associated to low-energy collective coordinates and so we only need to deal with frames that satisfy the decoherence condition to a very high degree of accuracy. It is worth emphasizing, though, that the Born-Markov property is not a necessary condition needed to define a frame but it is realistic and has the added bonus that it makes the stochastic dynamics of the conditioned state more tractable.

The decoherence condition means that the trajectory probabilities \eqref{tpr} are equal to the ensemble probabilities that follow from the density operator of the subsystem $\SS$, that is 
\EQ{
\rho\SUB{\SS}(t_n)=\Tr\SUB{\EE}\ket{\Psi(t_n)}\bra{\Psi(t_n)}=\sum_{\Ba^{(n)}}p_{\Ba^{(n)}}\ket{\Ba^{(n)}}\bra{\Ba^{(n)}}\SUB{\SS}\ ,
\label{zoz}
}
so in this case Born's rule has a dual meaning of ascribing probabilities to trajectories and ensembles.
This also means that a decoherent frame defines both an {\it unravelling\/}, to use the langauge of the theory of continuous measurement theory and quantum trajectories (described in appendix \ref{a1}) and a set of {\it consistent/decoherent histories\/} (described in appendix \ref{a2}).

It is important to note that the ensemble in \eqref{zoz} is {\it not\/} orthogonal because the pure states $\ket{\Ba^{(n)}}\SUB{\SS}$ need not be orthogonal, except for the last index as in \eqref{run}. In particular, it can be that $\ket{\Ba^{(n)}}\SUB{\SS}=\ket{\Bb^{(n)}}\SUB{\SS}$ for $\Ba^{(n)}\neq\Bb^{(n)}$.
But this could have been anticipated because intuitively the quantum system $\SS$ is not big enough to allow the set of states $\ket{\Ba^{(n)}}\SUB{\SS}$, labelled by trajectories, to be orthogonal; in general they will be massively over-complete. On the contrary, the environment $\EE$ has a much larger set of accessible states than $\SS$ and the states $\ket{\Ba^{(n)}}\SUB{\EE}$ can be expected to be orthogonal \eqref{kek}.

The conditioned state of $\SS$ includes a component that is the state in its own frame $\ket{\Ba^{(n)}}\SUB{\SS}$. It also includes the state of the complement of $\SS$, i.e.~the environment $\EE$, $\ket{\Ba^{(n)}}\SUB{\EE}$. The component state $\ket{\Ba^{(n)}}\SUB{\SS}$, the state of $\SS$ in its own frame, is the ``pointer state'' (the {\it real\/}, or ontic, state of $\SS$), whereas the state of the complement $\ket{\Ba^{(n)}}\SUB{\EE}$, in the frame $\SS$ is an ``epistemic state'', i.e.~contains the knowledge of how $\SS$ is correlated with the state of other subsystems of $\EE$ given the conditioned state of $\SS$ is $\ket{\Ba^{(n)}}\SUB{\SS}$:
\begin{equation*}
\text{In $\SS$'s frame:}\qquad\ket{\Psi(t_n)}=\sum_{\Ba^{(n)}}c_{\Ba^{(n)}}\cdots c_{\Ba^{(1)}}\ \overbrace{\underbrace{\ \ket{\Ba^{(n)}}\SUB{\SS}\ }_{\text{pointer state}}\quad\otimes~~\qquad\underbrace{\ket{\Ba^{(n)}}\SUB{\EE}}_{\text{epistemic}}}^{\text{conditioned state}}\qquad\qquad\phantom{.}
\end{equation*}
The fact that the pointer state of a subsystem is only a property of the subsystem frame itself and not any other frame, ensures that the formalism is local and causal. In passing, we note that there is much confusion in literature about whether the quantum state is to be regarded as ontic or epistemic: perhaps this is understandable because it is both.

\vspace{0.2cm}\noindent
{\bf Observer complementarity:\/} an important question is how do different frames relate to each other? More specifically, if the pointer state of a subsystem is the conditioned state $\ket{\Ba^{(n)}}\SUB{\SS}$ in its own frame, is this assignment to be thought of as a global fact, so true for ever other observer frame? This question boils down to what one observer (i.e.~frame) $\SS$ can say about the state of another subsystem frame $\SS'$. In general, $\SS$ only has the knowledge of the complement $\EE$, including $\SS'\subset\EE$, via the state $\ket{\Ba^{(n)}}\SUB{\EE}$. Is this knowledge enough to specify the conditioned state of $\SS'$? If $\SS$ and $\SS'$ are frames of macroscopic subsystems, then locality will ensure that they are entangled with different components of the joint environment $(\SS\cup\SS')^\perp$. In this case, the frames $\SS$ and $\SS'$ can be expected to consistently combine into the joint frame $\SS\cup\SS'$ since the state will have a decomposition of the form
\EQ{
\ket{\Psi(t_n)}=\sum_{\Ba^{(n)}\Bu^{(n)}}c_{\Ba^{(n)}\Bu^{(n)}}\cdots c_{\Ba^{(1)}\Bu^{(1)}}\ket{\Ba^{(n)}}\SUB{\SS}\otimes\ket{\Bu^{(n)}}\SUB{\SS'}\otimes \ket{\Ba^{(n)}\Bu^{(n)}}\SUB{(\SS\cup\SS')^\perp}\ ,
}
a form that manifests the consistency of the 3 frames $\SS$, $\SS'$ and $\SS\cup\SS'$. For macroscopic frames, we can expect all 3 frames $\SS$, $\SS'$ and $\SS\cup\SS'$ to be decoherent, meaning that the sets of states $\{\ket{\Ba^{(n)}}\SUB{\SS^\perp}\}$,  $\{\ket{\Bu^{(n)}}\SUB{\SS^{\prime\perp}}\}$ and $\{\ket{\Ba^{(n)}\Bu^{(n)}}\SUB{(\SS\cup\SS')^\perp}\}$ are orthogonal sets. Importantly, when the 3 frames fit together like this, it is consistent to define joint probabilities. These are simply the probabilities in the joint frame $p_{\Ba^{(n)}\Bu^{(n)}}$ such that
\EQ{
p_{\Ba^{(n)}}=\sum_{\Bu^{(n)}}p_{\Ba^{(n)}\Bu^{(n)}}\ ,\qquad p_{\Bu^{(n)}}=\sum_{\Ba^{(n)}}p_{\Ba^{(n)}\Bu^{(n)}}\ .
}

On the contrary, if the conditioned states of $\SS$ and $\SS'$ do not lift to the joint frame $\SS\cup\SS'$ then there is {\it observer complementarity\/} and no joint probabilities can be defined. We will see examples of this this when we discuss various gedankenexperiments in section \ref{s4}. To summarize, a global common classical reality is patched together by a set of consistent frames that lift to a joint global frame. We have already noted that it is to be expected that macroscopic frames {\it are\/} consistent because the subsystems will be entangled with causally separated parts of the environment.

\section{Classical localization}\label{s3}

In the last section, we have defined a set of trajectories of the collective coordinate subsystem $\SS$, defined by the states $\ket{\Ba^{(n)}}\SUB{\SS}$. The key question is what is the nature and dynamics of the conditioned state? In general, this problem would be formidable because it requires solving the interacting system $\SS\cup\EE$. Fortunately, the resulting dynamics has a high degree of universality and well-understood approximations can be made whilst still capturing the universal behaviour.

In order to put some flesh on the bones, let us consider the simplest macroscopic system $\SS$ where the only collective coordinate is the position $x$ in one dimension with a conjugate momentum $p$, effectively a particle, moving in a potential $V(x)$. In order to extract the universal dynamics, we can make a series of well understood and controllable approximations whose success relies on the universality of the problem. In particular, it does not really matter what type of environment we take, air at normal pressure, photons at room temperature, the Cosmic Microwave Background (CMB), etc. This kind of derivation has been performed and refined many times in the literature, for example, in the original work \cite{Joos:1984uk}, the book \cite{Sch} and the excellent article \cite{Hall}.

It is physically realistic to work in a limit where the velocity of the particle $\SS$ is much smaller than the velocity of the environmental particles, $p/M\ll k/m$. In this limit, an environmental particle reflects perfectly off the particle, the latter receiving a momentum kick of $2k$:
\begin{center}
\begin{tikzpicture}[scale=1]
\draw[fill=black] (0,0) circle (0.5cm);
\draw[fill=black!20] (-3.2,-0.1) circle (0.2cm);
\draw[very thick,->] (0,0)  -- (3,0);
\draw[very thick,dashed,->] (-3,-0.1)  -- (0,0)  -- (-2.5,0.1);
\node[right] at (3.3,0) {$p\to p+2k$};
\node[left] at (-3.7,0) {$k\to-k$};
\node at (0,1) {$\SS$};
\node at (-3.2,1) {$\EE$};
\node at (0,-1) {$M$};
\node at (-3.2,-1) {$m$};
\end{tikzpicture}
\end{center}

\noindent Hence, an initially non-entangled state becomes entangled:
\EQ{
\ket{\Psi}=\ket{\psi}\SUB{\SS}\otimes\int dk\,\phi(k)\ket{k}\SUB{\EE}\longrightarrow \int dk\,\phi(k)e^{2ikx/\hbar}\ket{\psi}\SUB{\SS}\otimes \ket{{-}k}\SUB{\EE}\ ,
}
where $\phi(k)$ is the momentum space wave function of the environment. In order to apply Born's rule to the final state, we calculate the density operator of $\SS$, $\rho\SUB{\SS}=\Tr\SUB{\EE}\ket{\Psi}\bra{\Psi}$. This effectively undergoes an impulsive change:
\EQ{
\rho\SUB{\SS}(t+\Delta t)=\int dk\,\big|\phi(k)\big|^2\,e^{2ikx/\hbar}\rho\SUB{\SS}(t) e^{-2ikx/\hbar}\ .
}
Realistically, the rate of scattering $\Gamma$ is large so that even over a microscopic time interval $\delta t$ many scattering events occur, $\delta t\gg\Delta t$;\footnote{It is worth pointing out here that because the effect of each scattering event is very weak it is consistent to take the time scale $\delta t$ large enough so that many events occur during the interval $\delta t$.} hence, over a time interval $\delta t$
\EQ{
\delta\rho\SUB{\SS}=\Gamma\delta t\int dk\,\big|\phi(k)\big|^2\,\Big(e^{2ikx/\hbar}\rho\SUB{\SS} e^{-2ikx/\hbar}-\rho\SUB{\SS}\Big)\ .
}
When the wavelength of the environmental particles is much larger than the spread of the state $\Delta x_\psi$, we can expand the exponentials to second order. Let us suppose that the environment has vanishing average momentum $\langle k\rangle\SUB{\EE}=\int dk\,k\big|\phi(k)\big|^2=0$, in which case:
\EQ{
\delta\rho\SUB{\SS}=\Lambda\big[x,\big[\rho\SUB{\SS},x\big]\big]\delta t\ ,\qquad \Lambda=\frac{2\Gamma\langle k^2\rangle\SUB{\EE}}{\hbar^2}\ ,
\label{art}
}
where $\langle k^2\rangle\SUB{\EE}=\int dk\,k^2\big|\phi(k)\big|^2$.
Including the self-Hamiltonian for evolution between the scattering events, gives the ``master equation'' for $\rho\SUB{\SS}$, 
\EQ{
\frac{\partial\rho\SUB{\SS}}{\partial t}=\frac1{i\hbar}\big[\frac{p^2}{2M}+V(x),\rho\SUB{\SS}\big]+\Lambda\big[\big[x,\rho\SUB{\SS}\big],x\big]\ .
\label{mst}
}

Now we can describe the evolution of the conditioned state $\ket{\psi}$ by writing the variation
\EQ{
\delta\rho\SUB{\SS}=\sum_{j=1}^Np_j\ket{\phi_j}\bra{\phi_j}-\rho\SUB{\SS}\ ,
}
for $\rho\SUB{\SS}=\ket{\psi}\bra{\psi}$, and matching to \eqref{mst}. This fixes $N{=}2$ and determines the states and probabilities:
\begin{center}
\begin{tikzpicture}[scale=0.7]
\draw[fill=black] (1,0) circle (0.3cm);
\draw[very thick] (-1,0)  -- (1,0);
\draw[very thick] (1,0)  -- (3,1);
\draw[-] (1,0)  -- (3,-1);
\node at (-2,0) {$\ket{\psi}$};
\node[right] at (3.5,1) {$\ket{\phi_2}=\big(1+H_\text{eff}\,\delta t/i\hbar\big)\ket{\psi}$};
\node[right] at (3.5,-1) {$\ket{\phi_1}=(x-\langle x\rangle_\psi)\ket{\psi}/\Delta x_\psi$};
\node[right] at (11.5,1) {$p_2=1-r\delta t$};
\node[right] at (11.5,-1) {$p_1=r\delta t$};
\end{tikzpicture}
\end{center}
where $\Delta x_\psi^2=\langle(x-\langle x\rangle_\psi)^2\rangle_\psi$ is the variance of the position and 
where the effective Hamiltonian is determined to be
\EQ{
H_\text{eff}=\frac{p^2}{2M}+V(x)-i\hbar\Lambda\big((x-\langle x\rangle_\psi)^2-\Delta x_\psi^2\big)\ .
\label{hef}
}
Note that this is both non-Hermitian and depends on the state $\ket{\psi}$ non-linearly. It is the non-Hermitian term that drives the localization of the state because the support of the state at $x\neq\langle x\rangle_\psi$ is suppressed exponentially.

The quantity $r$ is interpreted as the transition rate that the state $\ket{\psi}$ evolving according to the Schr\"odinger equation with Hamiltonian $H_\text{eff}$ makes a transition, or effectively a jump---although to be clear it is a microscopically small---into the orthogonal state $(x-\langle x\rangle_\psi)\ket{\psi}/\Delta x_\psi$. It is equal to
\EQ{
r=2\Lambda \Delta x_\psi^2\ .
\label{bub}
}
Note that $r/\Gamma$ is the probability that the state makes a transition during a single scattering event. This must be small for overall consistency.

There is an important subtle feature of the model that needs to be emphasized. Because the evolution of the density matrix of $\SS$ is captured by the stochastic dynamics of the conditioned state $\ket{\psi}$, it means that the condition \eqref{zoz} is satisfied and this implies that the decoherence condition \eqref{kek} is satisfied. So the fact that the dynamics of the subsystem $\SS$ can be expressed solely in terms of the density operator as in the master equation \eqref{mst}, which rests on the Born-Markov approximation, implies that the frame $\SS$ is decoherent. 

The universal dynamics of the conditioned state now reveals itself when realistic values of the parameters are considered. There is a separation of scales between the dynamics determined by the potential which is assumed to vary over classically macroscopic scales and the microscopic localization that is determined by a competition between the kinetic and non-Hermitian terms in $H_\text{eff}$. 

Localization occurs on microscopic scales and so in order to investigate it, we can set the potential to zero. A simple way to see the localization of a single wave packet is to solve \eqref{mst} (with $V=0$) using a (co-moving) harmonic oscillator basis, 
\EQ{
\psi(x,t)=\sum_{n=0}^\infty c_n\phi_{n}(x)e^{-i\omega(n+1/2)t}\ ,
\label{gss}
}
where $\phi_n(x)$ are the harmonic oscillator stationary states. The key point is that the frequency of the harmonic oscillator is complex
\EQ{
\omega=\sqrt{\frac{2\hbar\Lambda}{iM}}\ ,
\label{fre}
}
with $\IM\omega<0$, which means that the excited states in the sum \eqref{gss} decay relative to the ground state. The spatial extent of the latter is set by $\RE\omega$ and the normalizability of states is ensured since $\RE\omega>0$. The ground state 
is the pointer state \cite{SH}, an attractor for the dynamics of $H_\text{eff}$, a Gaussian state with position and momentum spreads of order
\EQ{
\Delta x_\text{p.s.}\thicksim \Big(\frac\hbar{M\Lambda}\Big)^{1/4}\ ,\qquad \Delta p_\text{p.s.}\thicksim (\hbar^3M\Lambda)^{1/4}\ .
\label{sup}
}
The approach to the attractor state, is exponential $\exp[-t/T_\text{loc.}]$ with a characteristic ``localization time''
\EQ{
T_\text{loc.}\thicksim \sqrt{\frac M{\hbar\Lambda}}\ .
\label{tlo}
}
We can relate this localization time to decoherence in the following way. If we ignore the self-Hamiltonian in \eqref{mst}, then the density operator in the position basis behaves as
\EQ{
\rho\SUB{\SS}(x,x')\thicksim \exp\big[-\Lambda(x-x')^2t\big]\ .
}
So the Lindblad terms have the effect of suppressing the off-diagonal components of the density operator in the position basis and we can define a decoherence time $1/\Lambda L^2$ for states with support on length scale $L$. This is, of course, what decoherence means. This effect is often called ``localization'' (e.g.~in \cite{Sch}) but this is potentially misleading because the state $\rho\SUB{\SS}$ is {\it not\/} localized in phase space, the diagonal components $\rho\SUB{\SS}(x,x)$ generally will be macroscopically spread out in phase space: decoherence is {\it not\/} localization. Rather it is the conditioned state $\ket{\psi}$ that is localized in phase space. However, the localization time \eqref{tlo} for the conditioned state is precisely the decoherence time $1/\Lambda L^2$, where the length scale $L$ is the spread of a pointer state $L=\Delta x_\text{p.s.}$.

It is interesting to consider the conditioned dynamics of a number of well separated pointer states $\ket{\psi}=\sum_i c_i\ket{\psi_i}$ each centred at $x_i$. 
We do this in detail in appendix \ref{a5}, following the analysis of \cite{SH}, where we show that localization picks out one of the components with probability given by Born's rule $|c_i|^2$. The localization time is determined by the scales $1/\Lambda|x_i-x_j|^2$. Given the large value of $\Lambda$, localization rapidly destroys a Schr\"odinger cat superposition state well before the components become macroscopically distinct.

\pgfdeclareimage[interpolate=true,width=12cm]{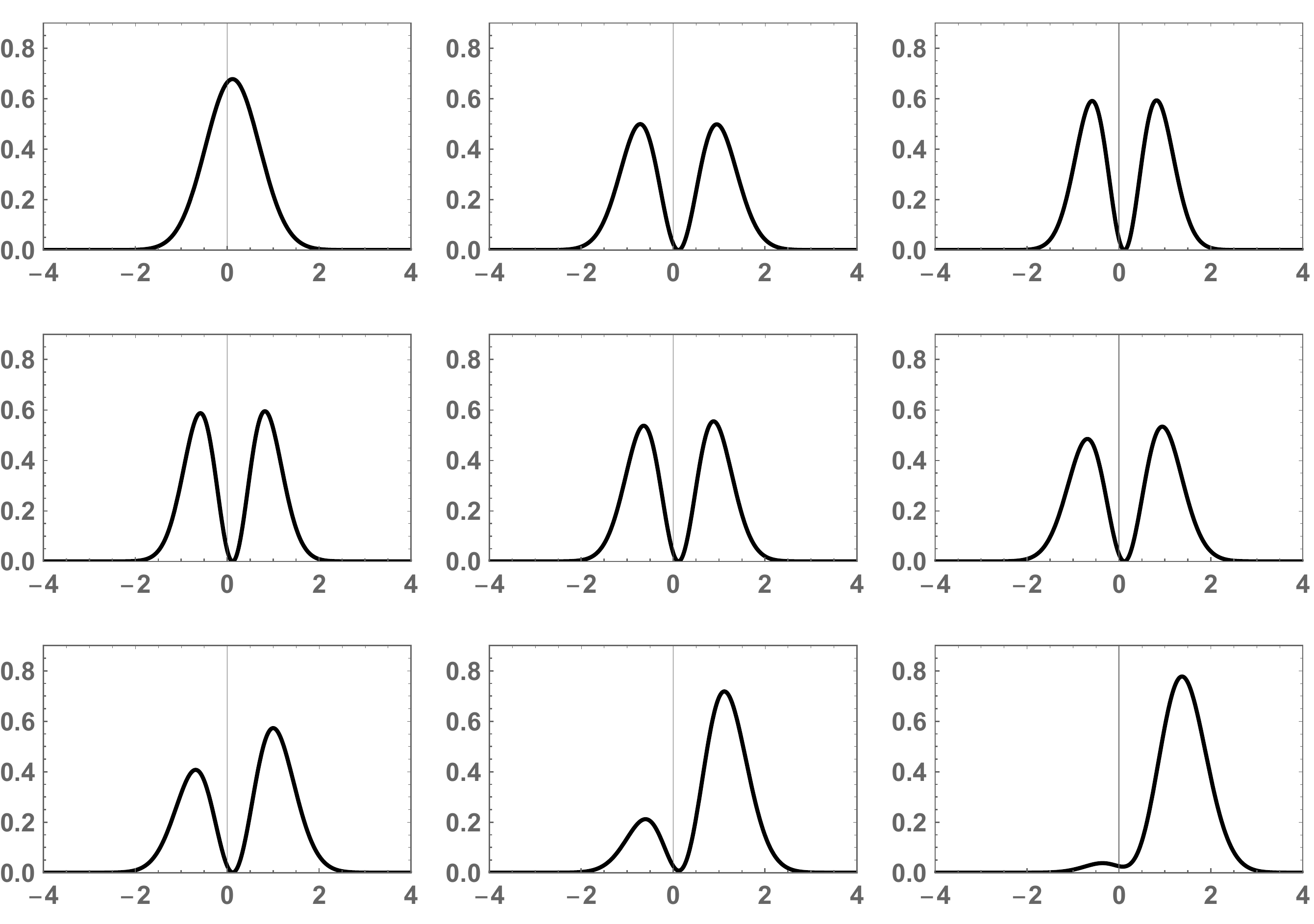}{f9}

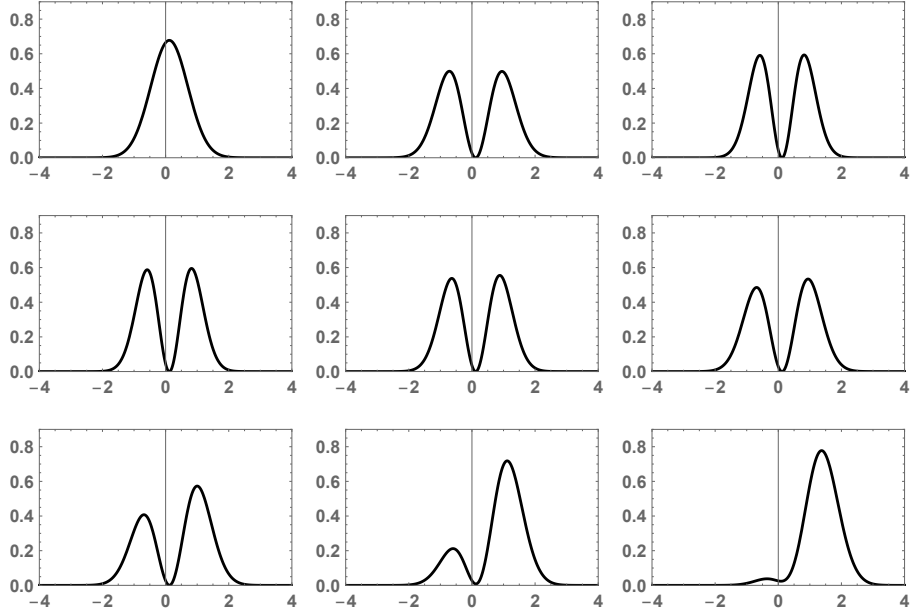
\begin{figure}
\begin{center}
\begin{tikzpicture}[scale=1]
\pgftext[at=\pgfpoint{0cm}{0cm},left,base]{\pgfuseimage{f9}} 
\end{tikzpicture}
\end{center}
\vspace{-0.5cm}
\caption{\footnotesize The effect of a jump on a near pointer state. The jump occurs between the first two plots. Over time, the state localizes onto one of the wave packets. The net effect is a jump sideways. Note further jumps are not considered here.}
\label{fig3}
\end{figure}

\vspace{0.2cm}\noindent
{\bf Effect of the jumps:\/} now let us consider the effect of the jumps on a single pointer state \cite{SH}. Since $\langle x\rangle_\psi$ lies within the support of a  single wave packet, the operator $J$ inserts a zero into the pointer state so that $J\ket{\psi}$ consists of two neighbouring wave packets. Then evolution by $H_\text{eff}$ has the effect of amplifying one of the new wave packets at the expense of the other: see figure \ref{fig3}.  The net effect is that the original wave packet is shifted sideways in phase space. Under the assumption that the initial wave packet is a pointer state, we can estimate the  shifts as 
\EQ{
\delta x\thicksim \pm\Big(\frac\hbar{M\Lambda}\Big)^{1/4}\ ,\qquad \delta p\thicksim \pm(\hbar^3M\Lambda)^{1/4}\ ,
\label{fur}
}
occurring with a rate
\EQ{
r\thicksim  \sqrt{\frac {\hbar\Lambda}M}\ .
\label{rux}
}
These jumps in phase space will be interpreted as Brownian motion at macroscopic scales. The jumps also ensure that the drift away from the classical trajectory induced by $H_\text{eff}$ is cancelled so that on the average, the conditioned state follows the classical trajectory. A simulation of a trajectory is shown in figure \ref{fig1}.

\pgfdeclareimage[interpolate=true,width=6cm]{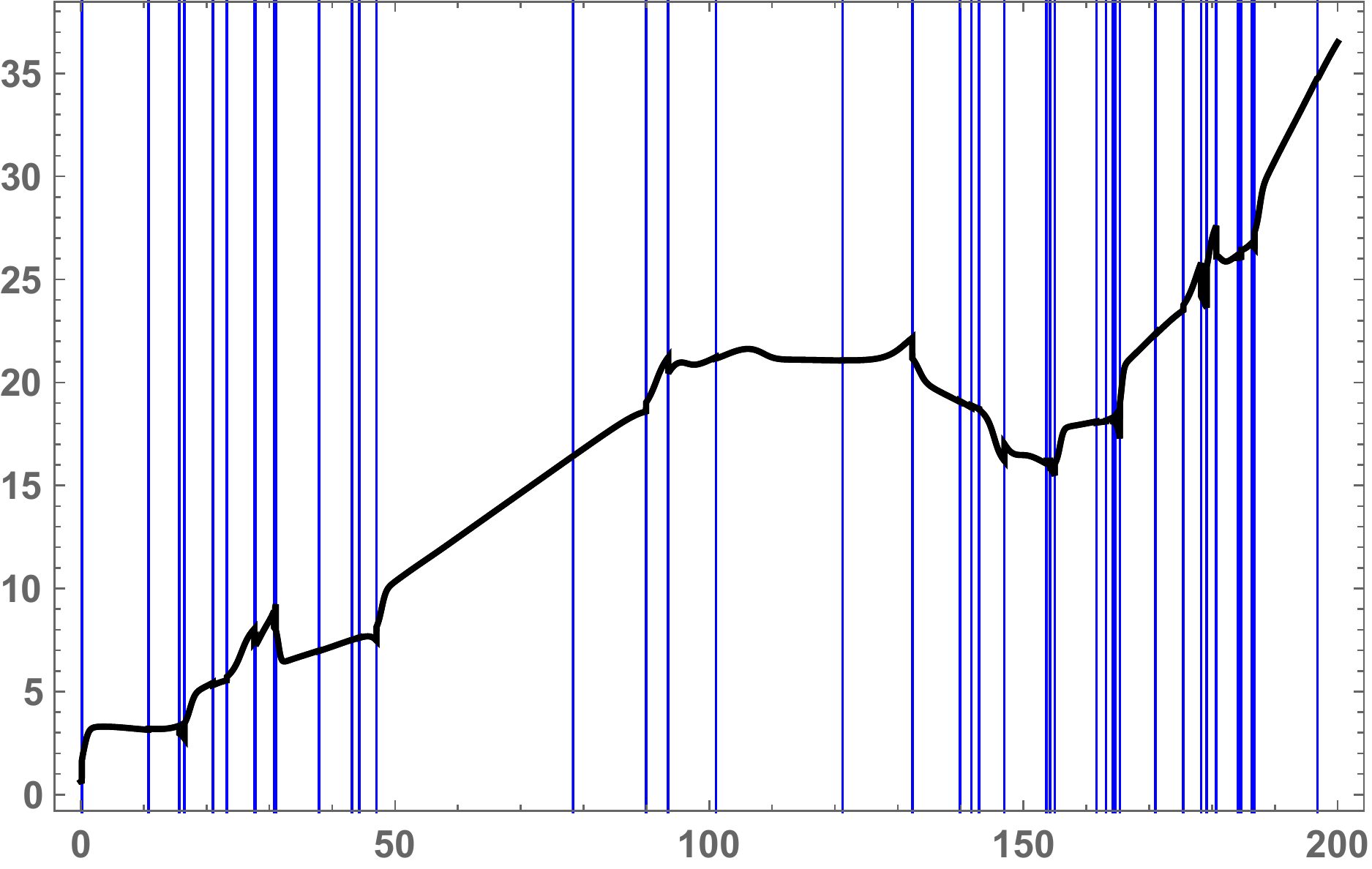}{f5}
\pgfdeclareimage[interpolate=true,width=6cm]{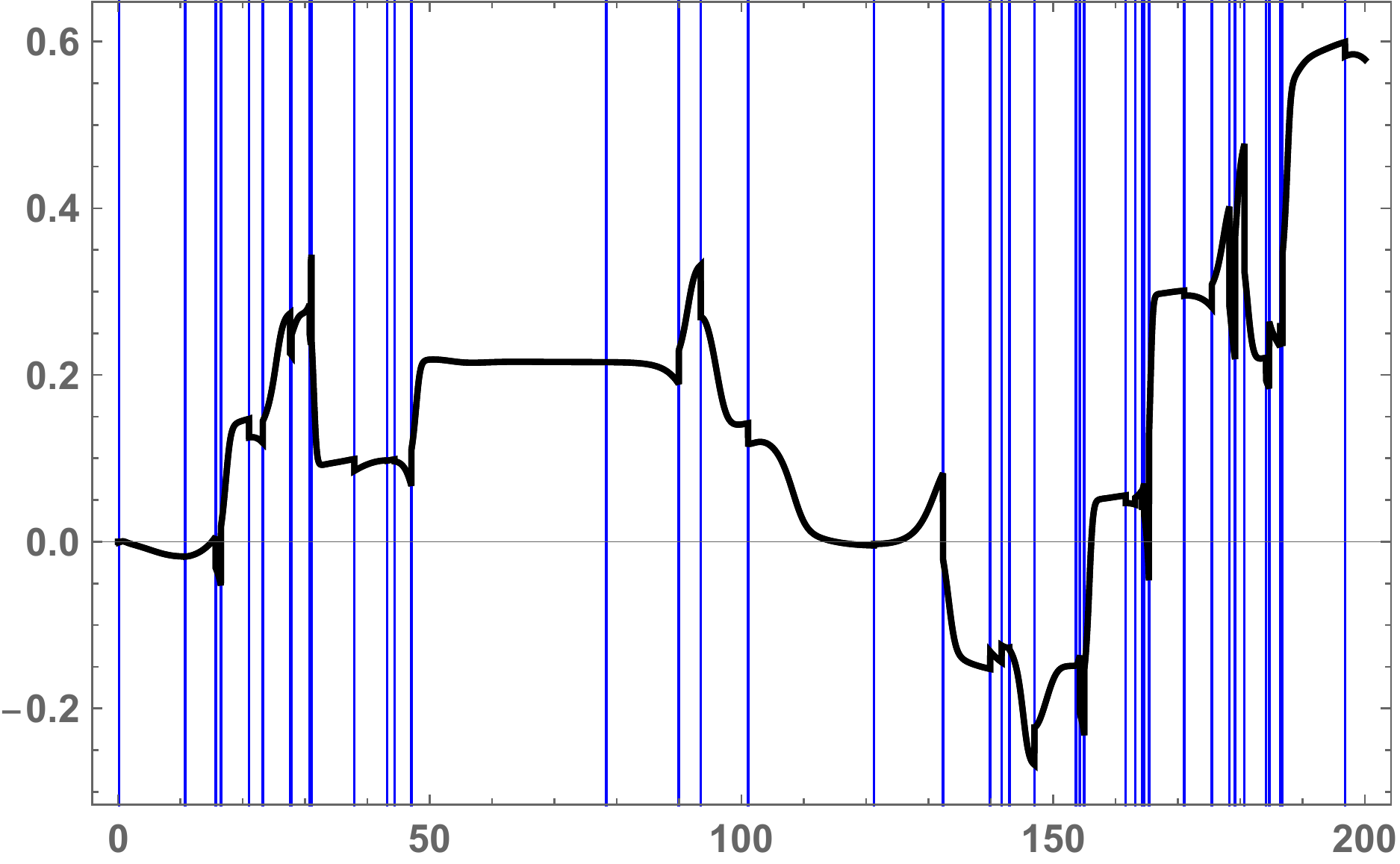}{f6}
\pgfdeclareimage[interpolate=true,width=6cm]{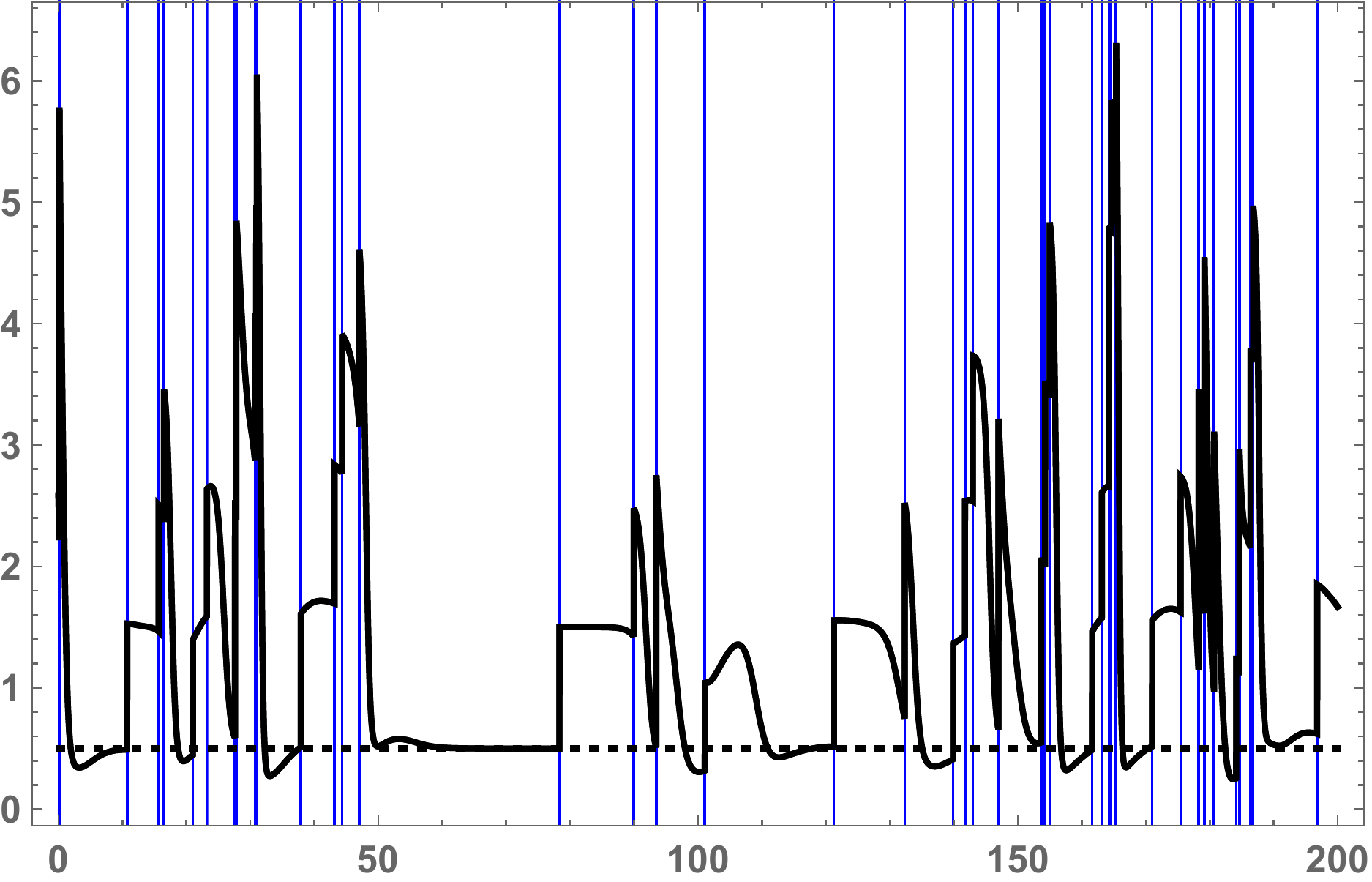}{f7}
\pgfdeclareimage[interpolate=true,width=6cm]{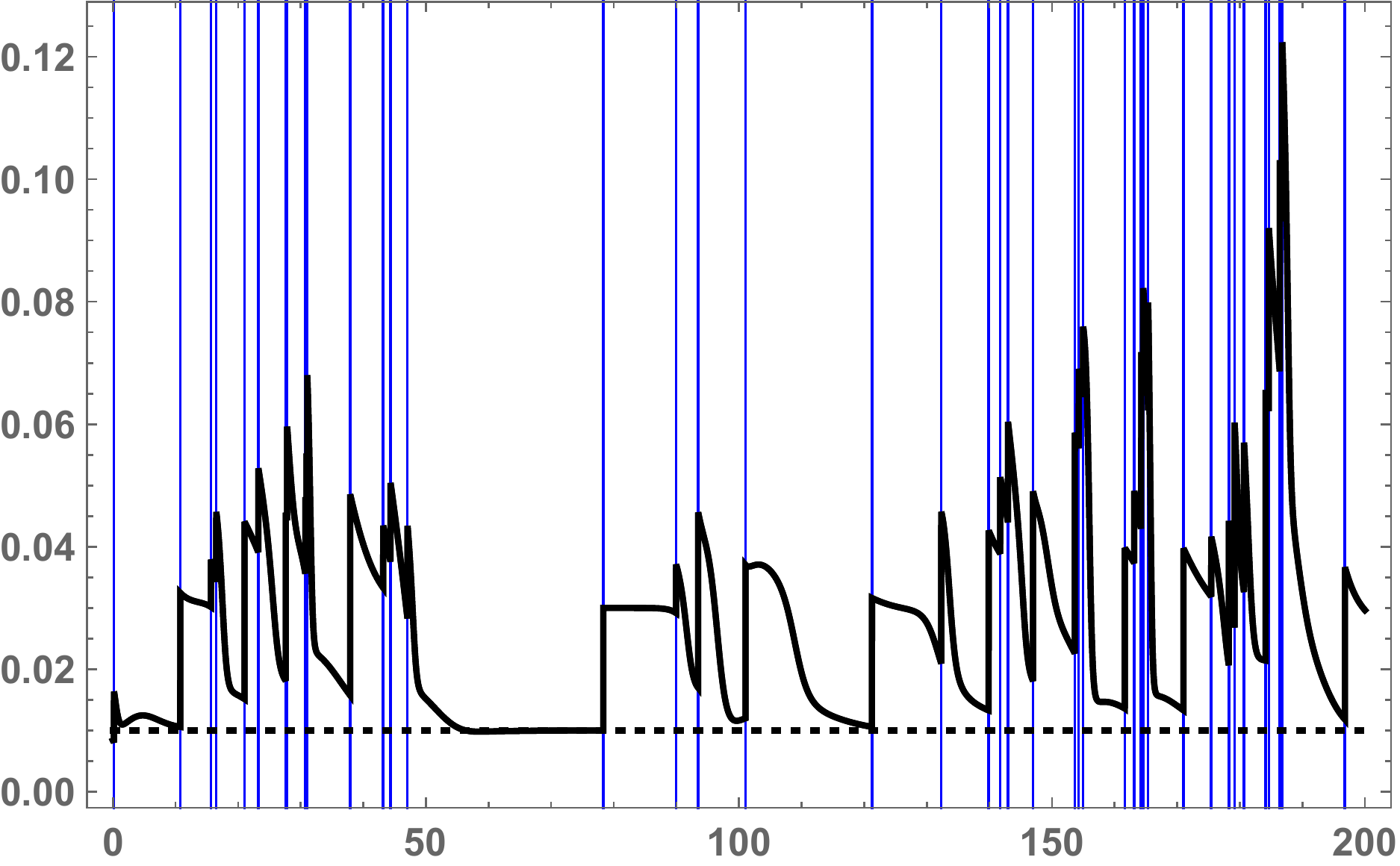}{f8}

\begin{figure}
\begin{center}
\begin{tikzpicture}[scale=1]
\node[left] at (0,2) {$\langle x\rangle_\psi$};
\node[left] at (8,2) {$\langle p\rangle_\psi$};
\node[left] at (0,-2.5) {$\Delta x^2_\psi$};
\node[left] at (8,-2.5) {$\Delta p^2_\psi$};
\pgftext[at=\pgfpoint{0cm}{0cm},left,base]{\pgfuseimage{f5}} 
\pgftext[at=\pgfpoint{8cm}{0cm},left,base]{\pgfuseimage{f6}} 
\pgftext[at=\pgfpoint{0cm}{-4.5cm},left,base]{\pgfuseimage{f7}} 
\pgftext[at=\pgfpoint{8cm}{-4.5cm},left,base]{\pgfuseimage{f8}} 
\end{tikzpicture}
\end{center}
\vspace{-0.5cm}
\caption{\footnotesize A simulation of the trajectory of the conditioned state for a free particle showing the phase space position and variances. The dotted lines show the variances of the pointer state and it is clear that the variances return to these values between jumps which are shown by the blue vertical lines. Note that the jumps occur in clusters because a jump increases $\Delta x_\psi^2$, and hence the rate of jumps, before localization ultimately sets in. The total number of jumps in the time interval shown is 44. The obvious noise and drift of the trajectory becomes Brownian motion for the macroscopic subsystem.}
\label{fig1}
\end{figure}
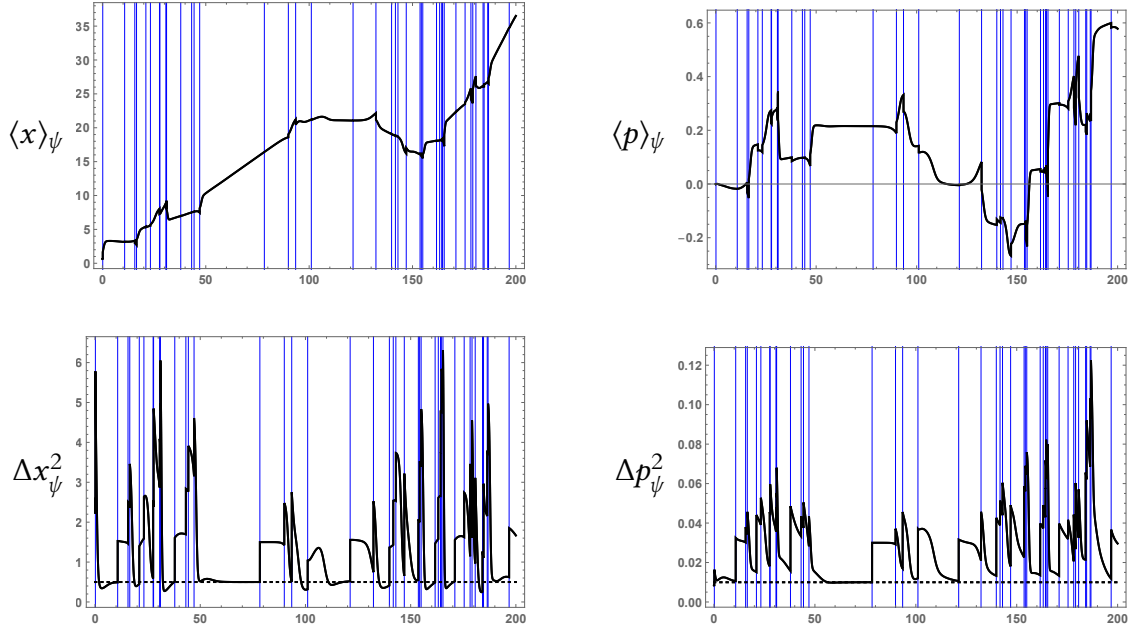

Let us estimate the scales involved for a dust particle of mass $1\,\text{g}$ and the values of $\Lambda$ estimated for three types of environment \cite{Sch}:
\begin{center}
\begin{tabular}{lcccc}
\toprule
Environment & $\Lambda$ (m${}^{-2}\,$s${}^{-1}$) & $\Delta x_\text{p.s.}$ (m) & $\Delta p_\text{p.s.}$ (kg$\,$m$\,$s${}^{-1}$) & $T_\text{loc.}$ (s)\\
\toprule
Air (atmos.~press.) & $10^{41}$ & $10^{-18}$ & $10^{-16}$ & $10^{-5}$\\
Photons (room temp.) & $10^{28}$ & $10^{-15}$ & $10^{-19}$ & $10^{2}$\\
CMB   & $10^{10}$ & $10^{-10}$ & $10^{-24}$ & $10^{10}$\\
\bottomrule
\end{tabular}
\end{center}
Of course, a realistic model should be three dimensional rather than one dimensional so our approach is admittedly crude, however, we believe it captures the universal behaviour irrespective of dimension.
The spreads of a pointer state $\Delta x_\text{p.s.}$ and $\Delta p_\text{p.s.}$, even for the CMB, are all safely microscopic, although the localization time scale is very slow in that case. The important implication is that localization is very efficient for realistic environments and at microscopic scales and, therefore, we can expect the state to be described by a microscopically narrow wave packet in phase space. 

\pgfdeclareimage[interpolate=true,width=12cm]{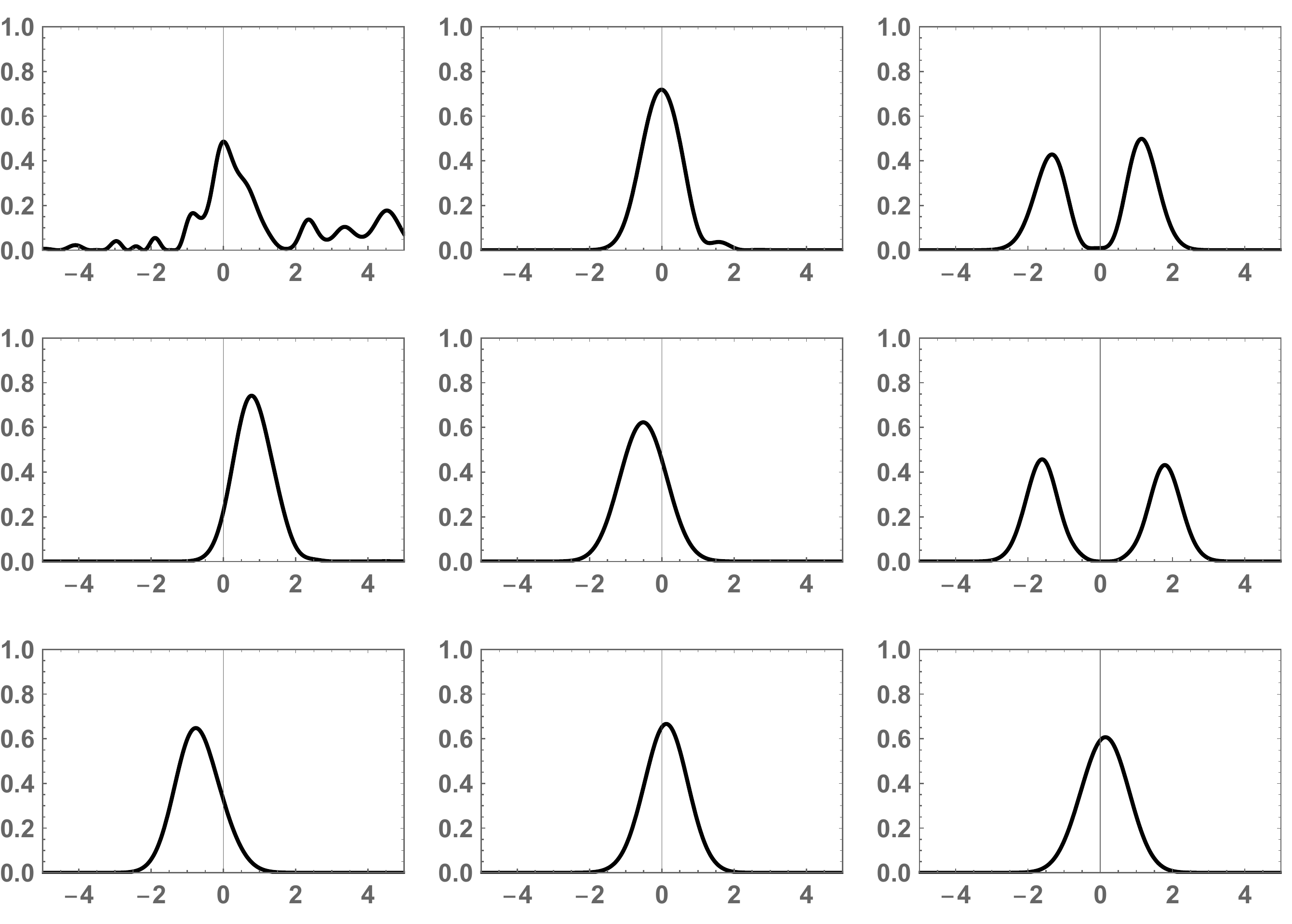}{f10}

\begin{figure}
\begin{center}
\begin{tikzpicture}[scale=1]
\pgftext[at=\pgfpoint{0cm}{0cm},left,base]{\pgfuseimage{f10}} 
\end{tikzpicture}
\end{center}
\vspace{-0.5cm}
\caption{\footnotesize Snapshots of the wave function for the trajectory in figure \ref{fig1}. The first plot is the initial chosen state. The state then localizes while at other instants it is caught after a jump as a double-peaked wave packet (causing the spikes in the variances in figure \ref{fig1}) which then localizes again. Note that the origin of $x$ has been shifted to approximately centre the wave packet at each snapshot.}
\label{fig4}
\end{figure}
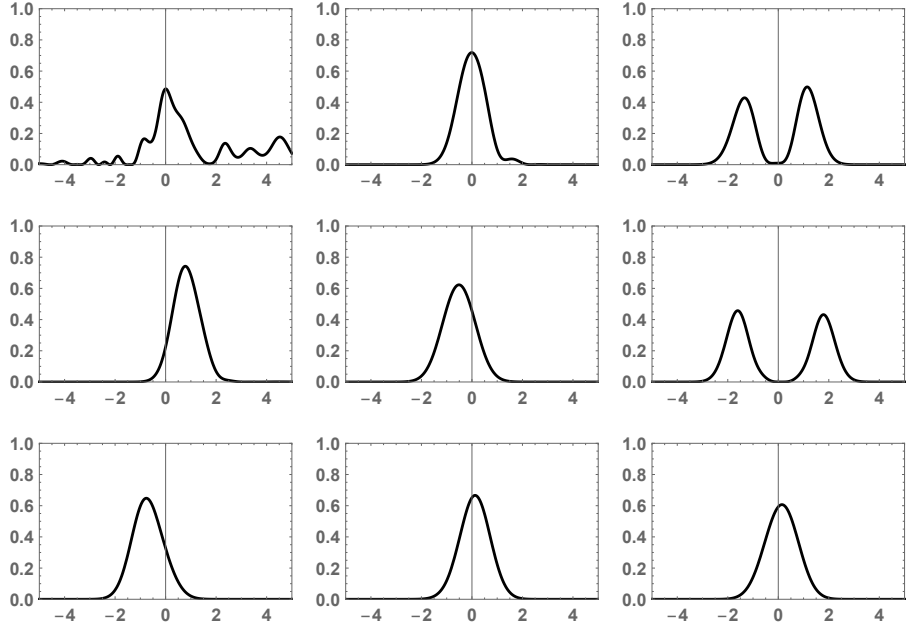

Consequently, on macroscopic scales we can describe the dynamics in terms of the motion of the wave packet $\bar x=\langle x\rangle_\psi$ and $\bar p=\langle p\rangle_\psi$ in phase space. At this point we can 
simply invoke Ehrenfest's Theorem to argue that the localized state follows a classical trajectory in phase space:
\EQ{
\frac{d\bar x}{dt}=\frac{\bar p}m\ ,\qquad \frac{d\bar p}{dt}=-\Big\langle\frac{dV(x)}{dx}\Big\rangle_\psi\approx-\frac{dV(\bar x)}{d\bar x}\ .
\label{tut}
}

Now we consider the effect of the random jumps and show how they emerge at the macroscopic level as small random corrections to Newton's equations that are precisely of the same order as the effective random jumps that a classical particle experiences coupled to a classical environment, i.e.~Brownian motion. We can estimate this as follows. Since $\Lambda\propto\hbar^{-2}$, it is the shifts \eqref{fur} in the momentum direction that are significant at macroscopic scales. Since the rate $r$ is large, the shifts by $\pm\delta p$ look like a random walk at macroscopic scales and Newton's equation \eqref{tut} becomes Langevin's equation
\EQ{
\frac{d\bar x}{dt}=\frac{\bar p}m\ ,\qquad \frac{d\bar p}{dt}=-\frac{dV(\bar x)}{d\bar x}+\sigma_p\,\xi\ ,
\label{lang}
}
where $\sigma_p\sim\delta p\sqrt{r}\sim\hbar\sqrt{\Lambda}={\mathscr O}(\hbar^0)$, where $\xi(t)$ is Gaussian noise with stochastic correlators
\EQ{
\mathscr E\big\{\xi(t)\big\}=0\ ,\qquad \mathscr E\big\{\xi(t)\xi(t')\big\}=\delta(t-t')\ .
}
This Langevin equation is precisely what one expects for classical Brownian motion, so the truly random quantum jumps manifest as the pseudo random jumps of the classical theory of Brownian motion. In order to see this, consider the  classical model where each scattering leads to an effective shift in the momentum of order
\EQ{
\delta p\thicksim\pm 2k\SUB{\EE}\ ,
}
with a rate $\Gamma$, where $k\SUB{\EE}$ is the characteristic momentum scale of a classical particle in the environment. This gives a random walk with $\sigma_p\sim k\SUB{\EE}\sqrt{\Gamma}$ which is identical to the quantum expression if we identify the classical scale $k\SUB{\EE}$ with the quantum expectation $\sqrt{\langle k^2\rangle_{\EE}}$. For a thermal environment $k\SUB{\EE}\sim\sqrt{mkT}$ and therefore we get the conventional relation of Brownian motion $\sigma_p\sim\sqrt{\gamma MkT}$, where $\gamma=\Gamma m/M$ is the relaxation/dissipation rate. Note that the more refined model described in \cite{SH} leads to the usual additional dissipation term in \eqref{lang}.

\vspace{0.2cm}\noindent
{\bf Chaotic instabilities:\/} a final issue to consider, is whether localization can be disrupted by chaos. In a classically chaotic system, trajectories can diverge in phase space exponentially $\sim\exp[\lambda t]$, where $\lambda$ is a local Lyapunov exponent. It is known that in the quantum analogue of such a classical system, the exponential divergence is mirrored in the exponential spreading of the (unconditioned) quantum state $\rho\SUB{\SS}$, at least up to the Ehrenfest time $T_\text{Ehr.}\sim\lambda^{-1}\log(S/\hbar)$, for the macroscopic action scale $S$ relevant to the system. At this time, a previously microscopic minimal uncertainty state becomes spread out on macroscopic scales and hence becomes completely non-classical.

A chaotic instability can be modelled locally in phase space by an upside down harmonic potential $V\sim -M\lambda^2x^2/2$. Using this simple potential, we can model the effect on the conditioned state. The addition of this potential modifies the frequency of the harmonic oscillator in \eqref{fre} to
\EQ{
\omega=\sqrt{-\lambda^2+2\hbar\Lambda/iM}\ .
}
In the harmonic oscillator basis, the excited states will decay because $\IM\omega<0$. The attractor state, the ground state, has a spread $\Delta x\sim\sqrt{\hbar/(M\RE\omega)}$. Localization requires that this scale is microscopically small which is ensured if $\lambda^2$ is not appreciably larger than $\hbar\Lambda/M$,
\EQ{
\lambda^2<\frac{2\hbar\Lambda}M\ ,\qquad\text{i.e.} \qquad  \lambda<1/T_\text{loc.}\ .
}

As an example, we can consider a chaotic system that is much discussed in the quantum chaotic literature, the chaotic intrinsic rotational motion of Saturn's moon Hyperion. We refer to Berry's excellent article for our estimates \cite{Berry}. Note that the system is angular, so $x$, $p$ and $M$ are replaced by the angle $\theta$, angular momentum $J$ and moment of inertia $I$. However, the previous expressions for the phase space spreads of a pointer state \eqref{sup} and the localization time \eqref{tlo} should still be valid with these replacements.

Hyperion's rotational motion has a Lyapunov exponent of about $\lambda\sim1/100\ \text{days}^{-1}$ and an Ehrenfest time of $\sim37\ \text{years}$. So in this alarmingly short time the unconditioned quantum state of the moon $\rho\SUB{\SS}$ becomes completely spread out on macroscopic scales. Of course, the unconditioned state is not the relevant state, rather it is the conditioned state that arises via the decoherence mechanism provided, for example, by the scattering of solar photons.\footnote{The use of solar photons as an environment is only one source of decoherence. Macroscopic bodies also have an internal environment which can act to decohere the collective coordinate subsystem. In this case, it is the scattering of phonons that leads to decoherence.} Berry roughly estimates the decoherence time for decoherence over the full angular scale $2\pi$ as $10^{-53}\ \text{s}$. Hence, the coupling $\Lambda\sim ((2\pi)^2\times10^{-53})^{-1}\sim 10^{51}\ \text{s}^{-1}$ (note the modified dimensions due to the angular system). We can estimate the moment of inertia via $I\sim MR^2$ with mass $M\sim5{\times}10^{18}\ \text{kg}$ and mean radius $R\sim135\ \text{km}$. From this data, we find that the pointer state of the rotational motion will have
\EQ{
\Delta\theta_\text{p.s.}\thicksim \Big(\frac\hbar{I\Lambda}\Big)^{1/4}\thicksim10^{-29}\ ,\qquad \Delta J_\text{p.s.}\thicksim (\hbar^3I\Lambda)^{1/4}\thicksim 10^{-6}\ \text{kg}\,\text{m}^2\,\text{s}^{-1}\ ,
\label{sup2}
}
compared with Hyperion's angular momentum $J\sim10^{24}\ \text{kg}\,\text{m}^2\,\text{s}^{-1}$. The localization time is
\EQ{
T_\text{loc.}\thicksim \sqrt{\frac I{\hbar\Lambda}}\thicksim 10\ \text{days}\ .
}
So in this case $\lambda T_\text{loc.}\sim1/10$ and localization of the quantum state is not disrupted by the chaotic instability.

Another application of this formalism is to the classicalization of the cosmological perturbations during inflation responsible for structure formation and the cosmic microwave background \cite{Hollowood:2018jgf}. In that case, it was found that localization occurs very efficiently during inflation and this explains how the inflationary fluctuations can be treated as essentially classical.

\section{Thought experiments}\label{s4}

In this section, we apply the formalism to three related thought experiments. In our vastly simplified treatment of these thought experiments, the goal is not to derive the classical behaviour of the measuring devices, that problem have already been solved and would require a much more detailed model, rather, it is to see what kind of picture of reality that arises from it. 

In order to apply the formalism, we will treat interactions between subsystems, e.g.~a qubit with a measuring device, as a single scattering event that leads to outcomes associated to the pointer basis of the device. A measuring device $\MM$ is taken to have an initial $\ket{m^0}$ and then a pair of pointer states $\ket{m^\pm}$ that indicate the outcome $\pm1$ of the measurement it performs. So we model $\MM$ as a qutrit. Implicitly we assume that the measuring device has an environment which provides decoherence for the frame of the measuring device. The implicit environment is usually important for the frame of the measuring device to satisfy the decoherence condition \eqref{kek}. So the state $\ket{m^i}$, for $i\in\{0,\pm\}$ is really $\ket{m^i}\ket{e^i}$ for orthogonal states of the environment $\ket{e^i}$.

\vspace{0.5cm}
\noindent{\bf (i)} In the first example, we will consider how the formalism describes the EPR experiment on an entangled qubit pair $\QQ_1$ and $\QQ_2$. In this case, separate environments for the measuring devices $\MM$ and $\NN$ are not needed because the qubits are sufficient by themselves to provide decoherence. To this end, we will take a Hilbert space in the form
\EQ{
\BH=\BH_{\MM}\otimes\BH_{\QQ_1}\otimes\BH_{\QQ_2}\otimes\BH_{\NN}\ .
}
In this section, we will leave the $\otimes$ implicit.

We take the initial state to be
\EQ{
\ket{\Psi(t_0)}=\frac1{\sqrt2}\ket{m^0}\big(\ket{z^+z^-}+\ket{z^-z^+}\big)\ket{n^0}\ .
\label{quz}
}
Here, the qubit states are eigenstates $\sigma_z\ket{z^\pm}=\pm\ket{z^\pm}$. The measuring devices are chosen to measure $\sigma_z$ on their qubit.
 
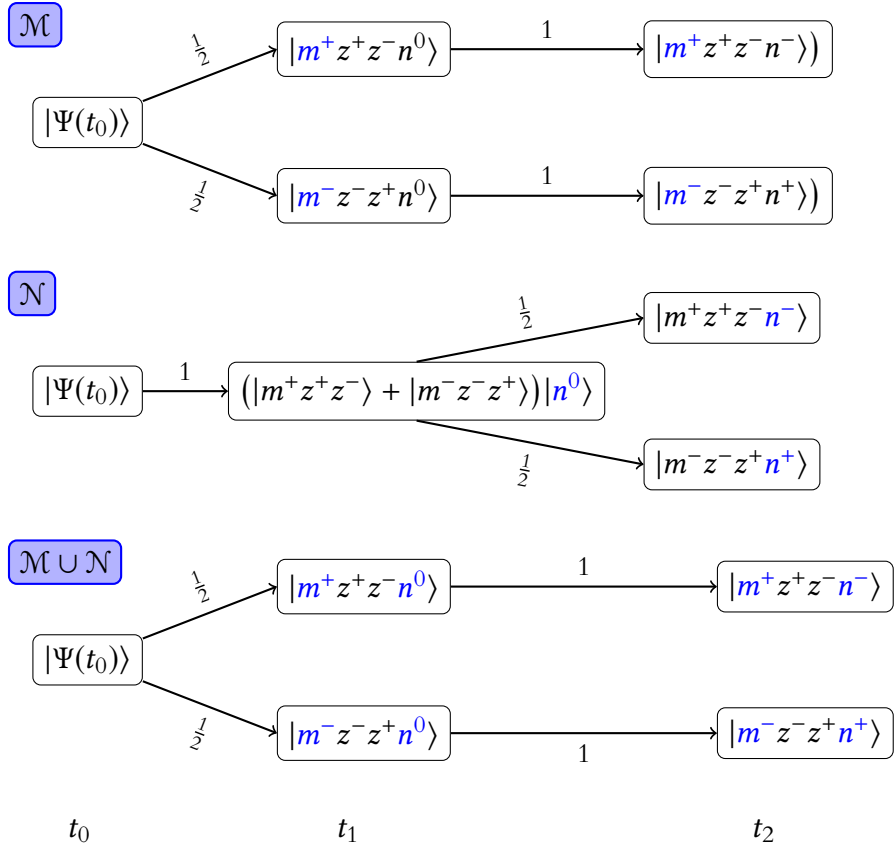
\begin{figure}
\begin{center}
\begin{tikzpicture}[scale=0.65]
\node[right,fill=blue!30, draw=blue, thick, rounded corners=3pt] (c1) at (-0.5,2) {$\MM$};
%\node[right] at (0.5,-4.5) {$t_0$};
%\node[right] at (6,-4.5) {$t_1$};
%\node[right] at (13.4,-4.5) {$t_2$};
\node[right,draw=black,rounded corners=3pt] (a1) at (0,0) {$\ket{\Psi(t_0)}$};
\node[right,draw=black,rounded corners=3pt] (a2) at (5,1.5) {$\ket{\BLUE{m^+}z^+z^-n^0}$};
\node[right,draw=black,rounded corners=3pt] (a3) at (5,-1.5) {$\ket{\BLUE{m^-}z^-z^+n^0}$};
\node[right,draw=black,rounded corners=3pt] (a4) at (12.5,1.5) {$\ket{\BLUE{m^+}z^+z^-n^-}\big)$};
\node[right,draw=black,rounded corners=3pt] (a5) at (12.5,-1.5) {$\ket{\BLUE{m^-}z^-z^+n^+}\big)$};
\draw[->,thick] (a1)  -- (a2.west) node[sloped,midway,above] {\footnotesize $\frac12$};
\draw[->,thick] (a1)  -- (a3.west) node[sloped,midway,below] {\footnotesize  $\frac12$};
\draw[->,thick] (a2)  -- (a4.west) node[sloped,pos=0.5,above] {\footnotesize  $1$};
\draw[->,thick] (a3)  -- (a5) node[sloped,midway,above] {\footnotesize  $1$};
\begin{scope}[yshift=-5.5cm]
\node[right,fill=blue!30, draw=blue, thick, rounded corners=3pt] (c1) at (-0.5,2) {$\NN$};
\node[right,draw=black,rounded corners=3pt] (a1) at (0,0) {$\ket{\Psi(t_0)}$};
\node[right,draw=black,rounded corners=3pt] (a2) at (4,0) {$\big(\ket{m^+z^+z^-}+\ket{m^-z^-z^+}\big) \ket{\BLUE{n^0}}$};
\node[right,draw=black,rounded corners=3pt] (a5) at (12.5,1.5) {$\ket{m^+z^+z^-\BLUE{n^-}}$};
\node[right,draw=black,rounded corners=3pt] (a6) at (12.5,-1.5) {$\ket{m^-z^-z^+\BLUE{n^+}}$};
\draw[->,thick] (a1)  -- (a2) node[sloped,midway,above] {\footnotesize $1$};
\draw[->,thick] (a2.north)  -- (a5.west) node[sloped,midway,above] {\footnotesize  $\frac12$};
\draw[->,thick] (a2.south)  -- (a6.west) node[sloped,pos=0.5,below] {\footnotesize $\frac12$};
\end{scope}
\begin{scope}[yshift=-11cm]
\node[right,fill=blue!30, draw=blue, thick, rounded corners=3pt] (c1) at (-0.5,2) {$\MM\cup\NN$};
\node[right] at (0.5,-3.5) {$t_0$};
\node[right] at (6,-3.5) {$t_1$};
\node[right] at (14.5,-3.5) {$t_2$};
\node[right,draw=black,rounded corners=3pt] (a1) at (0,0) {$\ket{\Psi(t_0)}$};
\node[right,draw=black,rounded corners=3pt] (a2) at (5,1.5) {$\ket{\BLUE{m^+}z^+z^-\BLUE{n^0}}$};
\node[right,draw=black,rounded corners=3pt] (a3) at (5,-1.5) {$\ket{\BLUE{m^-}z^-z^+\BLUE{n^0}}$};
\node[right,draw=black,rounded corners=3pt] (a4) at (14,1.5) {$\ket{\BLUE{m^+}z^+z^-\BLUE{n^-}}$};
%\node[right,draw=black,rounded corners=3pt] (a5) at (14,-1) {$\ket{m^-x^-x^+n^+}$};
%\node[right,draw=black,rounded corners=3pt] (a6) at (14,1) {$\ket{m^+x^+x^-n^-}$};
\node[right,draw=black,rounded corners=3pt] (a7) at (14,-1.5) {$\ket{\BLUE{m^-}z^-z^+\BLUE{n^+}}$}; 
\draw[->,thick] (a1)  -- (a2.west) node[sloped,midway,above] {\footnotesize $\frac12$};
\draw[->,thick] (a1)  -- (a3.west) node[sloped,midway,below] {\footnotesize  $\frac12$};
\draw[->,thick] (a2)  -- (a4.west) node[sloped,midway,above] {\footnotesize  $1$};
%\draw[->,thick] (a2)  -- (a6.west) node[sloped,pos=0.5,below] {\footnotesize  $\frac1{10}$};
%\draw[->,thick] (a3)  -- (a5.west) node[sloped,pos=0.5,above] {\footnotesize  $\frac12$};
\draw[->,thick] (a3)  -- (a7.west) node[sloped,pos=0.5,below] {\footnotesize  $1$};
\end{scope}
\end{tikzpicture}
\end{center}
\vspace{-0.5cm}
\caption{\footnotesize The possible trajectories of the conditioned state (non-normalized) in the three frames $\MM$, $\NN$ and $\MM\cup\NN$. The numbers above or below the lines give the probabilities for that transition. The pointer states (the conditioned state of a subsystem in its own frame) are shown in blue. It is clear that The 3 frames are all decoherent frames. It is also clear that the frames $\MM$ and $\NN$ lift consistently to the joint frame $\MM\cup\NN$ and so there is no observer complementarity in this case.}
\label{fig14}
\end{figure}

During the interval $[t_0,t_1]$, $\MM$ interacts with $\QQ_1$ so as to measure $\sigma_z$, leading to the state
\EQ{
\ket{\Psi(t_1)}=\frac1{\sqrt2}\big(\ket{m^+z^+z^-}+\ket{m^-z^-z^+}\big)\ket{n^0}\ .
}
Finally, during the interval $[t_1,t_2]$, $\NN$ interacts with $\QQ_2$ so as to measure $\sigma_z$, leading to the state
\EQ{
\ket{\Psi(t_2)}=\frac1{\sqrt{2}}\big(\ket{m^+z^+z^-n^-}+\ket{m^-z^-z^+n^+}\big)\ .
}

Now let us analyse the experiment from the point-of-view of subsystem frames. The trajectories and conditional probabilities of the two frames $\MM$ and $\NN$ are shown in figure \ref{fig14}. Also shown is the perspective of the joint frame $\MM\cup\NN$. Note that the joint frame is perfectly consistent with the individual frames: there is no observer complementarity. In particular, this means that it is meaningful to talk about the joint probabilities
\EQ{
p_{m^+n^-}=\frac12\ ,\qquad p_{m^-n^+}=\frac1{2}\ .
}
Note that in the subsystem frame formalism there are never any non-local effects. When $\MM$ measures $\sigma_z$ and the conditioned state changes from $\ket{\Psi(t_0)}$ to $\ket{m^+z^+z^-n^0}$, the state $\ket{m^+}$ is the actual state of $\MM$, a ontic/pointer state, whereas the state $\ket{z^+z^-n^0}$ of $\QQ_1\cup\QQ_2\cup\NN$ is only an epistemic state, giving the knowledge of how $\MM$ is correlated with $\QQ_1$, $\QQ_2$ and $\NN$.

\vspace{0.5cm}
\noindent{\bf (ii)} The second example is Wigner's friend thought experiment. It is interesting because it illustrates what happens when there is a breakdown of decoherence leading to observer complementarity. 

The set-up consists of a qubit $\QQ$ and 2 measuring devices, the friend $\FF$ and Wigner $\WW$ (with an implicit environment). Note that the qubit effectively decoheres states of $\FF$ and so an environment for $\FF$ is not needed so we will work in the Hilbert space 
\EQ{
\BH=\BH_{\FF}\otimes\BH_{\QQ}\otimes\BH_{\WW}\ .
}
The initial state is
\EQ{
\ket{\Psi(t_0)}=\ket{f^0}\big(\cos\phi\ket{\UP}+\sin\phi\ket{\DOWN}\big)\ket{w^0}\ .
}
We will assume, without loss of generality, that $\phi\in[0,\pi/4]$. During the interval $[t_0,t_1]$, the friend $\FF$ measures $\sigma_z$ of the qubit leading to the state
\EQ{
\ket{\Psi(t_1)}=\big(\cos\phi\ket{f^+\UP}+\sin\phi\ket{f^-\DOWN}\big)\ket{w^0}\ .
\label{pup}
}
So in $\FF$'s frame, there are 2 decoherent branches with probabilities $\cos^2\phi$ and $\sin^2\phi$. Note that the measurement is treated as a single scattering event where $\FF$, the system in this context, interacts with the qubit, which forms part of $\FF$'s environment $\FF^\perp=\QQ\cup\WW$. In order to simplify the notation, we will define the product states
\EQ{
\ket{F^\pm}\equiv\ket{f^\pm z^\pm}\ .
}

Then there is a second scattering event when $\WW$ measures $\sigma_x$ acting on the basis $\ket{F^\pm}$ of the combined system $\FF\cup\QQ$, i.e.
\EQ{
\sigma_x=\ket{F^+}\bra{F^-}+\ket{F^-}\bra{F^+}\ .
}
This measurement corresponds to the rotated basis
\EQ{
\frac1{\sqrt2}\big(\ket{F^+}+\ket{F^-}\big)\ ,\quad \frac1{\sqrt2}\big(\ket{F^+}-\ket{F^-}\big)\ .
\label{bas}
}  
It is at this point that the possibility of breaking the Born-Markov property and for the appearance of recoherence could occur because two components of $\FF$'s environment $\QQ$ and $\WW$ are not independent: $\WW$ is interacting with both $\FF$ and $\QQ$. Of course,  it would be practically impossible for $\WW$ to perform this kind of measurement in the real world with a macroscopic friend because it involves measuring in a basis of macroscopic superpositions of $\FF$'s state. 

After the measurement, the final state is
\EQ{
\ket{\Psi(t_2)}&=\frac{\cos\phi}2\,\boxed{\ket{F^+}\big(\ket{w^+}+\ket{w^-}\big)}+\frac{\cos\phi}2\,\boxed{\ket{F^-}\big(\ket{w^+}-\ket{w^-}\big)}\\[5pt]
&+\frac{\sin\phi}2\,\boxed{\ket{F^+}\big(\ket{w^+}-\ket{w^-}\big)}+\frac{\sin\phi}2\,\boxed{\ket{F^-}\big(\ket{w^+}+\ket{w^-}\big)}\ .
\label{pop}
}
In this expression, the first, respectively, second, line corresponds to unitary evolution of the component states of $\ket{\Psi(t_1)}$ in \eqref{pup}.

\begin{figure}
\begin{center}
\begin{tikzpicture}[scale=0.65]
\node[right,fill=blue!30, draw=blue, thick, rounded corners=3pt] (c1) at (-0.5,2) {$\FF$};
%\node[right] at (0.5,-4.5) {$t_0$};
%\node[right] at (6,-4.5) {$t_1$};
%\node[right] at (13.4,-4.5) {$t_2$};
\node[right,draw=black,rounded corners=3pt] (a1) at (0,0) {$\ket{\Psi(t_0)}$};
\node[right,draw=black,rounded corners=3pt] (a2) at (5.5,1.5) {$\ket{F^+ w^0}$};
\node[right,draw=black,rounded corners=3pt] (a3) at (5.5,-1.5) {$\ket{F^- w^0}$};
\node[right,draw=black,rounded corners=3pt] (a4) at (11.5,3) {$\ket{F^+}\big(\ket{w^+}+\ket{w^-}\big)$};
\node[right,draw=black,rounded corners=3pt] (a5) at (11.5,1) {$\ket{F^-}\big(\ket{w^+}-\ket{ w^-}\big)$};
\node[right,draw=black,rounded corners=3pt] (a6) at (11.5,-1) {$\ket{F^+}\big(\ket{w^+}-\ket{ w^-}\big)$};
\node[right,draw=black,rounded corners=3pt] (a7) at (11.5,-3) {$\ket{F^-}\big(\ket{w^+}+\ket{ w^-}\big)$}; 
\draw[->,thick] (a1)  -- (a2.west) node[sloped,midway,above] {\footnotesize $\cos^2\phi$};
\draw[->,thick] (a1)  -- (a3.west) node[sloped,midway,below] {\footnotesize  $\sin^2\phi$};
\draw[->,thick] (a2)  -- (a4.west) node[sloped,pos=0.5,above] {\footnotesize  $\frac12$};
\draw[->,thick] (a2)  -- (a5) node[sloped,midway,below] {\footnotesize  $\frac12$};
\draw[->,thick] (a3)  -- (a6) node[sloped,pos=0.5,above] {\footnotesize $\frac12$};
\draw[->,thick] (a3)  -- (a7.west) node[sloped,midway,below] {\footnotesize  $\frac12$};
\begin{scope}[yshift=-7cm]
\node[right,fill=blue!30, draw=blue, thick, rounded corners=3pt] (c1) at (-0.5,2) {$\WW$};
\node[right,draw=black,rounded corners=3pt] (a1) at (0,0) {$\ket{\Psi(t_0)}$};
\node[right,draw=black,rounded corners=3pt] (a2) at (4,0) {$\big(\cos\phi\ket{F^+}+\sin\phi\ket{F^-}\big) \ket{w^0}$};
\node[right,draw=black,rounded corners=3pt] (a5) at (11.5,1.5) {$\big(\ket{F^+}+\ket{F^-}\big)\ket{w^+}$};
\node[right,draw=black,rounded corners=3pt] (a6) at (11.5,-1.5) {$\big(\ket{F^+}-\ket{F^-}\big)\ket{ w^-}$};
\draw[->,thick] (a1)  -- (a2) node[sloped,midway,above] {\footnotesize $1$};
\draw[->,thick] (a2.north)  -- (a5.west) node[sloped,midway,above] {\footnotesize  $2a^2$};
\draw[->,thick] (a2.south)  -- (a6.west) node[sloped,pos=0.5,below] {\footnotesize $2b^2$};
\end{scope}
\begin{scope}[yshift=-14cm]
\node[right,fill=blue!30, draw=blue, thick, rounded corners=3pt] (c1) at (-0.5,2) {$\FF\cup\WW$};
\node[right] at (0.5,-4.5) {$t_0$};
\node[right] at (6,-4.5) {$t_1$};
\node[right] at (13.4,-4.5) {$t_2$};
\node[right,draw=black,rounded corners=3pt] (a1) at (0,0) {$\ket{\Psi(t_0)}$};
\node[right,draw=black,rounded corners=3pt] (a2) at (5.5,1.5) {$\ket{F^+ w^0}$};
\node[right,draw=black,rounded corners=3pt] (a3) at (5.5,-1.5) {$\ket{F^- w^0}$};
\node[right,draw=black,rounded corners=3pt] (a4) at (13,3) {$\ket{F^+ w^+}$};
\node[right,draw=black,rounded corners=3pt] (a5) at (13,-1) {$\ket{F^- w^+}$};
\node[right,draw=black,rounded corners=3pt] (a6) at (13,1) {$\ket{F^+ w^-}$};
\node[right,draw=black,rounded corners=3pt] (a7) at (13,-3) {$\ket{F^- w^-}$}; 
\draw[->,thick] (a1)  -- (a2.west) node[sloped,midway,above] {\footnotesize $\cos^2\phi$};
\draw[->,thick] (a1)  -- (a3.west) node[sloped,midway,below] {\footnotesize  $\sin^2\phi$};
\draw[->,thick] (a2)  -- (a4.west);
\draw[->,thick] (a2)  -- (a5);
%\draw[->,thick] (a3)  -- (a6);
%\draw[->,thick] (a3)  -- (a7.west);
%\draw[->,thick] (a3)  -- (a4);
\draw[->,thick] (a3)  -- (a4);
\draw[->,thick] (a3)  -- (a5);
\draw[->,thick] (a3)  -- (a6);
\draw[->,thick] (a3)  -- (a7.west);
\draw[->,thick] (a2)  -- (a6);
\draw[->,thick] (a2)  -- (a7);
%\node at (11,0) {{\Huge ?}}; 
%
\end{scope}
\end{tikzpicture}
\end{center}
\vspace{-0.5cm}
\caption{\footnotesize The possible trajectories of the conditioned state (non-normalized) in the frames $\FF$, $\WW$ and $\FF\cup\WW$. The numbers above or below the lines give the probabilities for that transition (in the joint case between $t_1$ and $t_2$ they are all $\tfrac14$). It is clear that the frames $\FF$ and $\WW$ do not lift consistently to $\FF\cup\WW$ and so there is observer complementarity in this case. Note that $\FF$ and $\WW$ are decoherent frames while $\FF\cup\WW$ exhibits recoherence.}
\label{fig8}
\end{figure}
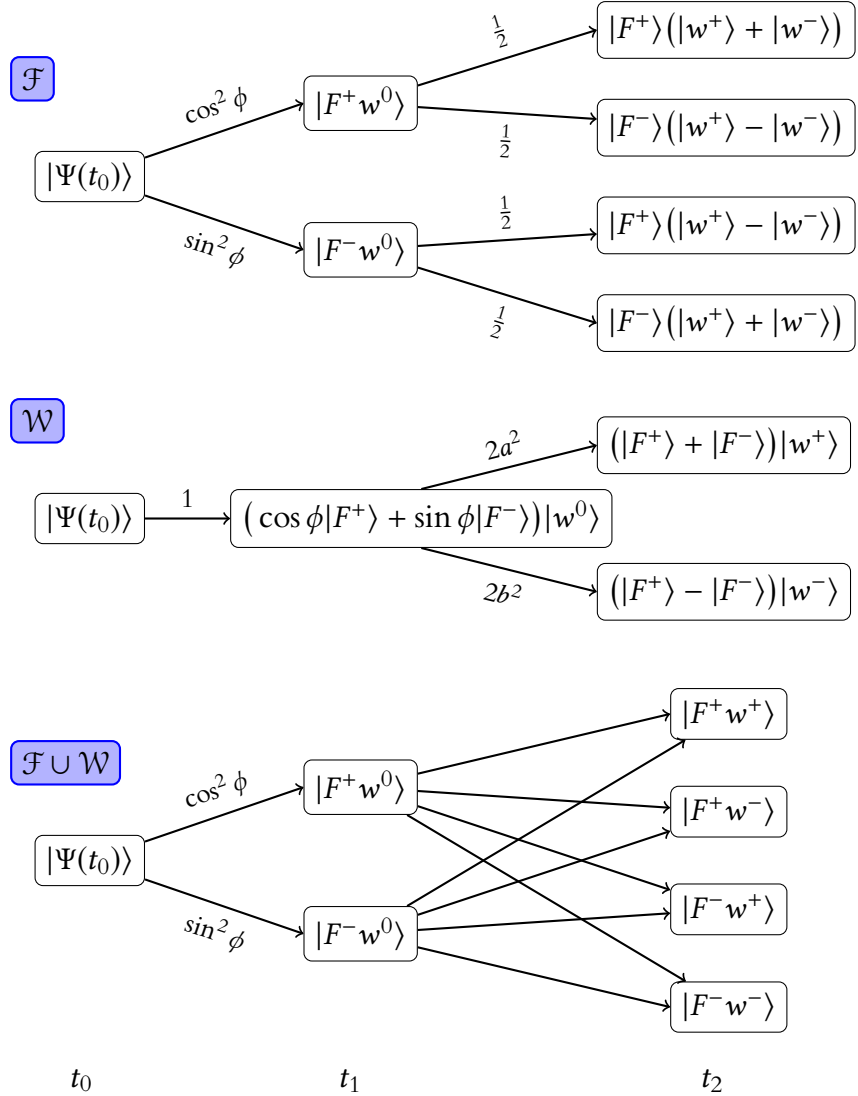

The terms in the boxes in \eqref{pop} correspond to the possible conditioned state in $\FF$'s frame. Note that the states of the environment $\FF^\perp=\QQ\cup\WW$ are all orthogonal and so the decoherence condition \eqref{kek} is satisfied. We can also write the state in terms of the conditioned states in $\WW$'s frame:
\EQ{
\ket{\Psi(t_2)}=a\,\boxed{\big(\ket{F^+}+\ket{F^-}\big) \ket{w^+}}+b\,\boxed{\big(\ket{F^+}-\ket{F^-}\big)\ket{ w^-}}\ ,
\label{pop3}
}
where we have defined
\EQ{
a=\frac{\cos\phi+\sin\phi}2\ ,\qquad b=\frac{\cos\phi-\sin\phi}2\ .
}
Note that $\WW$ also defines a decoherent frame. The conditioned states and trajectories for the frames $\FF$, $\WW$ and the joint frame $\FF\cup\WW$, are shown in figure \ref{fig8}. 

The question is whether there is observer complementarity? Can the frames $\FF$ and $\WW$ be lifted consistently into a joint frame $\FF\cup\WW$? $\FF$ can reason about $\WW$'s outcome by using the conditioned states in $\FF$'s frame: see figure \ref{fig8}. $\FF$'s conclusion is that joint probabilities for the four possible joint states $\ket{F^+w^+}$, $\ket{F^+w^-}$, $\ket{F^-w^+}$ and $\ket{F^-w^-}$ are all equal to $\frac14$.
However, $\WW$ can reason about the joint state in his frame. In that case, the joint states $\ket{F^\pm w^+}$ have probability $(\cos\phi+\sin\phi)^2/4$ while $\ket{F^\pm w^-}$ have probability $(\cos\phi-\sin\phi)^2/4$.
The extreme case occurs when $\phi=\pi/4$. In that case, in $\WW$'s frame, the outcome $\WW=-1$ never occurs whereas $\FF$ would reason that it occurs with probability $\frac12$. The mismatch between the joint probabilities assigned by the two frame here, is an indication that there are no consistent joint probabilities for $\FF$ and $\WW$ and there {\it is\/} observer complementarity.

In to order to investigate further, we need to consider the joint frame $\FF\cup\WW$.
After the first measurement at $t_1$ there are 2 conditioned states of $\FF\cup\WW$ which match those in $\FF$'s frame. The non-trivial part of the story occurs at the next time step. If we evolve the 2 conditioned states at $t_1$ to $t_2$ then we would identify the conditioned states of $\FF\cup\WW$'s as the 8 states that appear in the decompositions
\EQ{
&U(t_2,t_1)\ket{F^+ w^0}=\frac12\,\boxed{\ket{F^+ w^+}}+
\frac12\,\boxed{\ket{F^+ w^-}}+
\frac12\,\boxed{\ket{F^- w^+}}-
\frac12\,\boxed{\ket{F^- w^-}}\ ,\\[5pt]
&U(t_2,t_1)\ket{F^- w^0}=\frac12\,\boxed{\ket{F^+ w^+}}-
\frac12\,\boxed{\ket{F^+ w^-}}+
\frac12\,\boxed{\ket{F^- w^+}}+
\frac12\,\boxed{\ket{F^- w^-}}\ .
\label{br1}
}
But we can see that the conditioned states in the two branches \eqref{br1} are not decoherent because the corresponding states of the environment, i.e.~$\QQ$ and the implicit environment of $\WW$,\footnote{This means that states $\ket{w^\pm}$ are really $\ket{w^\pm e^\pm}$, for environmental states $\ket{e^\pm}$.} in the 2 branches are not orthogonal. In fact, we could say that the branches are maximally recoherent because the 2 sets of 4 states are actually equal. The conclusion is that in this case there is observer complementarity that is caused by the fact that the joint frame $\FF\cup\WW$ violates the decoherence condition. In particular, one cannot define consistently joint probabilities for $\FF$ and $\WW$. 

Of course the problem is that we cannot exhibit the observer complementarity in a real experiment where both the frames $\FF$ and $\WW$ are macroscopic. However, by doubling up the number of qubits, friends and Wigners, $\QQ_i$, $\FF_i$ and $\WW_i$, with a suitable 2-qubit initial state, one can engineer experiments that can manifest the complementarity. The general idea is that the existence of joint probabilities $p_{\FF_1\WW_1\FF_2\WW_2}$ are incompatible with quantum mechanical predictions of various joint measurements that can be performed on the enlarged system. There are 2 variations of this doubled set up that we consider in (iii) and (iv) below.

\vspace{0.2cm}
\noindent {\bf(iii)} CHSH version: the complementarity described above can be detected by a Bell inequality type test \cite{Br1,Br2} by choosing a suitable initial state of the qubits. If we denote the observables as $\sigma_x^{(i)}$ and $\sigma_z^{(i)}$ for $\FF_i\cup\QQ_i$, then it is simple to show that existence of a joint probability distribution $p_{\FF_1\FF_2\WW_1\WW_2}$ implies a Clauser-Horne-Shimony-Holt inequality
\EQ{
\big|\langle\sigma^{(1)}_z\sigma^{(2)}_z\rangle+\langle\sigma^{(1)}_z\sigma^{(2)}_x\rangle+\langle\sigma^{(1)}_x\sigma^{(2)}_z\rangle-\langle\sigma^{(1)}_x\sigma^{(2)}_x\rangle\big|\leq2\ .
}
However, quantum mechanics can violate the inequality and does so maximally when the initial state of the qubits is taken as 
\EQ{
\frac1{\sqrt2}\big(\cos\theta\ket{z^+z^+}+\sin\theta\ket{z^+z^-}+\sin\theta\ket{z^-z^+}-\cos\theta\ket{z^-z^-}\big)\ ,
}
with $\theta=\pi/8$ when the left-hand side of the inequality is $2\sqrt2$.

Remarkably this test has been performed in a real experiment \cite{PP} and the quantum mechanical violation was observed showing that observer complementarity is a real phenomena. 

\vspace{0.5cm}
\noindent{\bf (iv)} Finally, we consider the doubled-up variation of the Wigner's friend thought experiment described by Frauchiger and Renner \cite{FR1}.\footnote{The title of \cite{FR1} is misleading. It is not that quantum mechanics cannot describe the consistent use of itself, it is simply that quantum mechanics does not, in general, allow different frames to be embedded in a consistent whole.} 

There are 2 qubits $\QQ_i$, 2 friends $\FF_i$ and 2 Wigners $\WW_i$, $i=,1,2$. Friend $\FF_i$ measures $\sigma_z$ of qubit $\QQ_i$ and then Wigner $\WW_i$ measures $\sigma_x$ of the combined system $\FF_i\cup\QQ_i$ corresponding to the basis \eqref{bas}. We will take the measurements by $\FF_i$ to occur in the interval $[t_0,t_1]$ and the measurements by $\WW_i$ in $[t_1,t_2]$. The initial state of the qubits is
\EQ{
\frac1{\sqrt3}\big(\ket{z_1^+z_2^+}+\ket{z_1^+z_2^-}+\ket{z_1^-z_2^-}\big)\ .
\label{quz3}
}

We can now make the following quantum mechanical predictions for various joint measurements:
\begin{enumerate} 
\item As is evident from \eqref{quz3}, for a measurement of $\FF_1$ and $\FF_2$ the outcome $(-1,+1)$ has vanishing probability $p_{\FF_1=-1,\FF_2=+1}=0$.
\item For a measurement of $\FF_1$ and $\WW_2$, the outcome $\WW_2=-1$ implies $\FF_1=-1$. We can see this from the decomposition of the qubit state of \eqref{quz3}
\EQ{
\frac1{\sqrt{12}}(2\ket{z_1^+}+\ket{z_1^-}\big)\underbrace{\big(\ket{z_2^+}+\ket{z_2^-}\big)}_{\WW_2=+1}-\frac1{\sqrt{12}}\overbrace{\ket{z_1^-}}^{\FF_1=-1}\underbrace{\big(\ket{z_2^+}-\ket{z_2^-}\big)}_{\WW_2=-1}\ .
\label{quz4}
}
\item For a measurement of $\WW_1$ and $\FF_2$, the outcome $\WW_1=-1$ implies $\FF_2=+1$. We can see this from the decomposition of the qubit state of \eqref{quz3}
\EQ{
\frac1{\sqrt{12}}\underbrace{\big(\ket{z_1^+}-\ket{z_1^-}\big)}_{\WW_1=-1}\overbrace{\ket{z^+_2}}^{\FF_2=+1}
+\frac1{\sqrt{12}}\underbrace{\big(\ket{z_1^+}+\ket{z_1^-}\big)}_{\WW_1=+1}(\ket{z_2^+}+2\ket{z_2^-}\big)\ .
\label{quz5}
}
\item For a measurement of $\WW_1$ and $\WW_2$, the outcome $(-1,-1)$ has probability $p_{\WW_1=-1,\WW_2=-1}=\frac1{12}$. We can see this from another decomposition of the qubit state of \eqref{quz3}
\EQ{
&\frac3{\sqrt{48}}\overbrace{\big(\ket{z_1^+}+\ket{z_1^-}\big)}^{\WW_1=+1}\underbrace{\big(\ket{z_2^+}+\ket{z_2^-}\big)}_{\WW_2=+1}-\frac1{\sqrt{48}}\overbrace{\big(\ket{z_1^+}+\ket{z_1^-}\big)}^{\WW_1=+1}\underbrace{\big(\ket{z_2^+}-\ket{z_2^-}\big)}_{\WW_2=-1}\\ 
&+\frac1{\sqrt{48}}\overbrace{\big(\ket{z_1^+}-\ket{z_1^-}\big)}^{\WW_1=-1}\underbrace{\big(\ket{z_2^+}+\ket{z_2^-}\big)}_{\WW_2=+1}+\frac1{\sqrt{48}}\overbrace{\big(\ket{z_1^+}-\ket{z_1^-}\big)}^{\WW_1=-1}\underbrace{\big(\ket{z_2^+}-\ket{z_2^-}\big)}_{\WW_2=-1}\ .
\label{quz6} 
}
\end{enumerate}

These statements are mutually incompatible if there exists joint probabilities $p_{\FF_1\WW_1\FF_2\WW_2}$. For instance, (2) says that outcome $\WW_2=-1$ implies $\FF_1=-1$ and (3) that outcome $\WW_1=-1$ implies $\FF_2=+1$. Point (4) says that this occurs with probability $\frac1{12}$, but point (1) says that the joint probability for $\FF_1=-1$ and $\FF_2=+1$ vanishes.

That fact that joint probabilities $p_{\FF_1\WW_1\FF_2\WW_2}$ cannot be defined is because there is observer complementarity and 
the frames $\FF_1$, $\FF_2$, $\WW_1$ and $\WW_2$ cannot all fit together consistently in a single joint frame. In particular, the joint frame breaks the decoherence condition \eqref{kek} at $t_2$. To see this, we note that at $t=t_1$ the joint frame has 3 conditioned states corresponding to
\EQ{
\ket{\Psi(t_1)}=\frac1{\sqrt3}\boxed{\ket{F_1^+F_2^+w_1^0w_2^0}}+\frac1{\sqrt3}\boxed{\ket{F_1^+F_2^-w_1^0w_2^0}}+\frac1{\sqrt3}\boxed{\ket{F_1^-F_2^-w_1^0w_2^0}}\ .
\label{quz10}
}
These 3 branches are decoherent once one remembers that $\ket{F^\pm_i}=\ket{f^\pm_i}\ket{z^\pm_i}$ and the state $\ket{w^0_i}$ include an implicit environment. But at $t=t_2$ each of these 3 branches spawns 16 conditioned states corresponding to $\ket{F_1^{\epsilon_1}F_2^{\epsilon_2}w_1^{\epsilon_3}w_2^{\epsilon_4}}$, for $\epsilon_i\in\{\pm\}$. Since the same set of 16 is spawned by each of the 3 branches at $t_1$ there is maximal recoherence and the decoherence condition \eqref{kek} is broken. This leads to observer complementarity and the fact that joint probabilities cannot be defined. 

\vspace{0.5cm}
\begin{center}
*****
\end{center}
\vspace{0.5cm}

We have demonstrated that ``standard'' quantum mechanics in the form of Schr\"odinger's Equation and Born's rule can predict the localization of states in phase space and classical mechanics for the localized state of a macroscopic system immersed in a generic environment. It is important that the approach is built on the locality and causality of microscopic physics in that the notion of a local frame of reference plays a key r\^ole. The fact that one can predict the emergence of classical physics at macroscopic scales within standard quantum theory makes any discussion of ``interpretations'' of quantum mechanics unnecessary.

\appendix
\appendixpage

\section{Unravelling}\label{a1}

The theory of quantum trajectories is concerned with the dynamics of the density operator $\rho\SUB{\SS}$ of a subsystem like $\SS$ conditioned on measurements made on its environment (for a nice introduction containing original references, see \cite{JacobsSteck}). In this formalism, $\SS$ is usually a microscopic system, e.g.~an individual atom whose environment $\EE$ is the electromagnetic field. The trajectories are defined by a series of measurements (often taken in a limit to be continuous) that are made on $\EE$ by an external measuring device. These measurements define a series of projection operators acting on $\BH_{\EE}$. This should be contrasted with the present application, where the subsystem $\SS$ is taken to macroscopic not microscopic and where there are a series of projection operators on $\BH_{\EE}$ but with no actual measurements being made on $\EE$.

If the subsystem is initially in the state $\rho(t_0)=\ket{\Ba^{(0)}}\bra{\Ba^{(0)}}\SUB{\SS}$, then an unravelling corresponds to a decomposition 
\EQ{
\rho\SUB{\SS}(t)={\mathscr E}\big\{\ket{\Ba(t)}\bra{\Ba(t)}\SUB{\SS}\big\}\ ,
}
where $a(t)$ are a set of stochastic ``trajectories'' of some auxiliary stochastic variables and ${\mathscr E}$ is a stochastic average. Importantly there can be many consistent unravellings because there is no requirement of orthogonality on the states $\ket{\Ba(t)}\SUB{\SS}$. 

It is clear that, in our case, the states $\ket{\Ba^{(n)}}\SUB{\SS}$ provide a discrete unravelling of this type, where the stochastic trajectories are the strings $\Ba^{(n)} =(a_1,\ldots,a_n)$ and the associated trajectory of pure states of $\SS$:
\EQ{
\ket{\Ba^{(1)}}\SUB{\SS} \ \longrightarrow\ \ket{\Ba^{(2)}}\SUB{\SS} \ \longrightarrow \ \cdots\ \longrightarrow \ \ket{\Ba^{(n)}}\SUB{\SS}
}
and where the stochastic average is defined as 
\EQ{
\mathscr E\big\{\cdots\big\}=\sum_{\Ba^{(n)}}p_{\Ba^{(n)}}\big\{\cdots\big\}\ .
}
Note that the fact that we have an unravelling is rather trivial in this context because the trajectories are encoded directly in the states at a given time.

In the language of quantum trajectories, the unravelling corresponds to a measurement scheme defined by the projectors $\mathbb P^\EE_{\Ba^{(n)}}=\ket{\Ba^{(n)}}\bra{\Ba^{(n)}}\SUB{\EE}$ on $\BH_{\EE}$. Of course, in the present context, there is no actual external measuring device. The continuum limit of the unravelling that we are led to by applying Born's rule for each microscopic scattering was first formulated by Di\'osi \cite{Diosi1,Diosi2}.

\section{Consistent/decoherent histories}\label{a2}

The formalism of consistent \cite{RG,RO}, or decoherent \cite{GH}, histories attempts to identify when it is possible to associate probabilities consistently to trajectories. This will only be possible when the trajectories are suitably decoherent. Unfortunately in this formalism, like the unravellings in appendix \ref{a1}, there is no uniqueness: there are many consistent histories.

In the approach of consistent histories, a history (in the Schr\"odinger Picture) in the time interval $[t_0,t_n]$ is associated to a chain of projection operators and unitaries, 
\EQ{
\mathbb C_{\Ba^{(n)} }=\mathbb P_{\Ba^{(n)}}U(t_n,t_{n-1})\mathbb P_{\Ba^{(n-1)}}U(t_{n-1},t_{n-2})\cdots U(t_2,t_1)\mathbb P_{\Ba^{(1)}}U(t_1,t_0)\ ,
}
with the completeness relations at each $p=1,2,\ldots,n$:
\EQ{
\sum_{\Ba^{(p)}}\mathbb P_{\Ba^{(p)}}=1\ .
}
In order to apply this formalism to the present approach, we need to use the branch dependent histories where the projectors at a given time depend on the previous states of the branch \cite{GellMann:1995cu}, i.e.~the trajectory $\Ba^{(n)}$.

The condition that the histories do not interfere can be expressed as a condition on the decoherence functional:
\EQ{
\mathscr D(\Ba^{(n)} ,\Bb^{(n)}  )=\bra{\Psi(t_0)}\mathbb C_{\Ba^{(n)} }^\dagger\mathbb C_{\Bb^{(n)}  }\ket{\Psi(t_0)}=p_{\Ba^{(n)}}\delta_{\Ba^{(n)}\Bb^{(n)}}\ .
\label{xzz}
}
In our context, we have
\EQ{
\mathbb P_{\Ba^{(n)}}=\ket{\Psi_{\Ba^{(n)}}}\bra{\Psi_{\Ba^{(n)}}}\ ,
}
where $\ket{\Psi_{\Ba^{(n)}}}=\ket{\Ba^{(n)}}\SUB{\SS}\otimes\ket{\Ba^{(n)}}\SUB{\EE}$, 
and then the decoherence condition \eqref{xzz} follows from the fact that
\EQ{
\mathbb C_{\Ba^{(n)} }\ket{\Psi(t_0)}=c_{\Ba^{(n)}}c_{\Ba^{(n-1)}}\cdots c_{\Ba^{(1)}}\ket{\Psi_{\Ba^{(n)} }}\ ,
}
and the orthogonality of the states $\ket{\Psi_{\Ba^{(n)} }}$, following from the orthogonality \eqref{kek} of the states $\ket{\Ba^{(n)}}\SUB{\EE}$ of the environment, with respect to the whole history $\Ba^{(n)}$. Therefore, the decoherence condition \eqref{xzz} is then equivalent to the orthogonality condition \eqref{kek}.

In fact, the histories satisfy a more general decoherence condition \cite{GellMann:1995cu}
\EQ{
\bra{\Psi(t_0)}\mathbb C_{\Ba^{(n)}}^\dagger {\EuScript O}\SUB{\SS}\mathbb C_{\Bb^{(n)}}\ket{\Psi(t_0)}=p_{\Ba^{(n)}}\bra{\Ba^{(n)}}{\EuScript O}\SUB{\SS}\ket{\Ba^{(n)}}\SUB{\SS}\delta_{\Ba^{(n)}\Bb^{(n)}}\ ,
\label{xzz2}
}
where ${\EuScript O}\SUB{\SS}$ is {\it any\/} operator acting on the Hilbert space $\BH_\SS$.

It is worth remarking that the existence of consistent histories relies on the fact that the states $\ket{\Ba^{(n)}}\SUB{\EE}$ are orthogonal as in \eqref{kek} but not on the orthogonality of the states $\ket{\Ba^{(n)}}\SUB{\SS}$. However, when the states $\ket{\Ba^{(n)}}\SUB{\SS}$ are orthogonal in their last index, as implied by Born's rule \eqref{run}, we can then write the histories in a way that is completely intrinsic to $\SS$ \cite{Paz:1993tg}. To this end, the initial density operator of $\SS$ is
\EQ{
\rho\SUB{\SS}(t_0)=\ket{\Ba^{(0)}}\bra{\Ba^{(0)}}\SUB{\SS}\ .
}
where $\Ba^{(0)}=\emptyset$.
At the next time step, after a single scattering event, the density operator can be written
\EQ{
\rho\SUB{\SS}(t_1)=\sum_{a_1}K_{\Ba^{(1)}}\rho\SUB{\SS}(t_0)K_{\Ba^{(1)}}^\dagger\ ,
}
for a set of Krauss operators $K_{\Ba^{(1)}}$. This is a way of writing the evolution in terms of a quantum channel.
The evolution then continues in this way:
\EQ{
\rho\SUB{\SS}(t_n)=\sum_{\Ba^{(n)}}K_{\Ba^{(n)}}\cdots K_{\Ba^{(2)}}K_{\Ba^{(1)}}\rho\SUB{\SS}(t_0)K_{\Ba^{(1)}}^\dagger K_{\Ba^{(2)}}^\dagger \cdots K_{\Ba^{(n)}}^\dagger 
\ .
}
The Krauss operators are defined in terms of the unitarity evolution operator of the total system $\SS\cup\EE$ as
\EQ{
K_{\Ba^{(n)}}=\bra{\Ba^{(n)}}U(t_n,t_{n-1})\ket{\Ba^{(n-1)}}\SUB{\EE}
}
and they satisfy the completeness relation
\EQ{
\sum_{a_n}K_{\Ba^{(n)}}^\dagger K_{\Ba^{(n)}}=1\ .
}
Note the sum here is only over the last index of $\Ba^{(n)}$. We can write the evolution of the conditioned state using the Krauss operators as
\EQ{
\ket{\Ba^{(n)}}\SUB{\SS}=\frac1{c_{\Ba^{(n)}}}K_{\Ba^{(n)}}\ket{\Ba^{(n-1)}}\SUB{\SS}\ .
}

Note that the density operator of $\SS$ can be written as
\EQ{
\rho\SUB{\SS}(t_n)=\sum_{\Ba^{(n)}}\Tr\SUB{\EE}\big(\mathbb C_{\Ba^{(n)}}\ket{\Psi(t_0)}\bra{\Psi(t_0)}\mathbb C_{\Ba^{(n)}}^\dagger\big)
}
and the decoherence functional can be written in a way that is intrinsic to $\SS$:
\EQ{
\mathscr D(\Ba^{(n)},\Bb^{(n)})=\sum_{\Bc^{(n)}}\Tr\SUB{\SS}\big(\mathbb P^{(\SS)}  _{\Ba^{(n)}}K_{\Bc^{(n)}}\cdots
\mathbb P^{(\SS)}  _{\Ba^{(1)}} K_{\Bc^{(1)}}\rho\SUB{\SS}(t_0)K_{\Bc^{(1)}}^\dagger\mathbb P^{(\SS)}  _{\Bb^{(1)}}\cdots K_{\Bc^{(n)}}^\dagger\mathbb P^{(\SS)}  _{\Bb^{(n)}}\big)\ ,
}
where we have defined the projectors operators on $\SS$
\EQ{
\mathbb P^{(\SS)} _{\Ba^{(n)}}=\ket{\Ba^{(n)}}\bra{\Ba^{(n)}}\SUB{\SS}\ .
}
Note that the projectors are orthogonal on the last index due to the orthogonality condition \eqref{run}; for $\Ba^{(n)}=(a_1,\ldots,a_{n-1},a_n)$ and $\Ba^{\prime(n)}=(a_1,\ldots,a_{n-1},a'_n)$
\EQ{
\mathbb P^{(\SS)} _{\Ba^{(n)}}\mathbb P^{(\SS)} _{\Ba^{\prime(n)}}=\delta_{a_na'_n}\mathbb P^{(\SS)} _{\Ba^{(n)}}\ .
}

\section{General master equation and conditioned dynamics}\label{a5}

In this appendix we consider the inverse problem of going from some known master equation for the dynamics of the density operator $\rho\SUB{\SS}$ to dynamics of the conditioned state. This is know as the process of {\it unravelling a master equation\/}.

The most general dynamical equation for a density operator $\rho\SUB{\SS}$ consistent with positivity takes the form
\EQ{
\frac{\partial\rho\SUB{\SS}}{\partial t}=\frac1{i\hbar}[H,\rho\SUB{\SS}]+\frac12\sum_{\mu\nu}r_{\mu\nu}\big(2F_\mu\rho\SUB{\SS} F_\nu^\dagger-F_\nu^\dagger F_\mu\rho\SUB{\SS}-\rho\SUB{\SS} F_\nu^\dagger F_\mu\big)\ ,
\label{urr}
}
where $F_\mu$ are a basis of operators on $\BH_{\SS}$ normalized via $\Tr\SUB{\SS}(F_\mu^\dagger F_\nu)=\delta_{\mu\nu}$. In the case when $\BH_\SS$ has finite dimension $N$, $\mu,\nu=1,2,\ldots,N^2$. In \eqref{urr}, $r_{\mu\nu}$ is a Hermitian matrix with non-negative eigenvalues. 

Now if $\rho\SUB{\SS}=\ket{\psi}\bra{\psi}$, where $\ket{\psi}$ is the instantaneous conditioned state, then we can choose a basis that is adapted to $\ket{\psi}$ in which (say) $\ket{\psi}=(0,\ldots,0,1)$ and the operators can be taken as $E_{ab}$, a matrix with 1 in position $(a,b)$. There are $N$ operators that do not annihilate $\ket{\psi}$ which we will define as $F_j\equiv E_{jN}$, $j=1,\ldots,N-1$ and $F_N\equiv E_{NN}$. The other $N(N-1)$ operators $E_{ab}$, $b\neq N$, that annihilate $\ket{\psi}$ will be denoted collectively as $F_{\hat\mu}$.

Hence, if $\rho\SUB{\SS}=\ket{\psi}\bra{\psi}$ at $t$ then the variation at $t+\delta t$ is
\EQ{
\delta\rho\SUB{\SS}&=\frac1{i\hbar}[H,\rho\SUB{\SS}]\delta t-\frac12\sum_{\hat\mu j}\big(r_{j\hat\mu}F_{\hat\mu}^\dagger F_j\rho\SUB{\SS}+r_{\hat\mu j}\rho\SUB{\SS} F_j^\dagger F_{\hat\mu}\big)\delta t\\ &+ \frac12\sum_j[r_{jN}F_j-r_{Nj}F_j^\dagger,\rho\SUB{\SS}]\delta t+\frac12\sum_{ij}r_{ij}\big(2F_i\rho\SUB{\SS} F_j^\dagger-F_j^\dagger F_i\rho\SUB{\SS}-\rho\SUB{\SS} F_j^\dagger F_i\big)\delta t\ .
\label{urr2}
}

The matrix with elements $r_{ij}$, $i,j=1,2,\ldots,N-1$, can be diagonalized by a unitary transformation, $UrU^\dagger=\text{diag}(r_j)$, with $r_j$ real and positive. Let us define rotated operators by $F_i=\sum_jU_{ij}J_j/\sqrt{r_j}$, for which 
\EQ{
\langle J_j\rangle_\psi=0\ ,\qquad\langle J_i^\dagger J_j\rangle_\psi=r_i\delta_{ij}\ ,
\label{gut}
}
where we have defined $\langle \cdots\rangle_\psi\equiv\bra{\psi}\cdots\ket{\psi}$. In terms of these, the final term in \eqref{urr2} takes the form
\EQ{
\frac12\sum_{j=1}^{N-1}\big(2J_j\rho\SUB{\SS} J_j^\dagger-J_j^\dagger J_j\rho\SUB{\SS}-\rho\SUB{\SS} J_j^\dagger J_j\big)\delta t\ ,
}
so this brings \eqref{urr2} into the form
\EQ{
\delta\rho\SUB{\SS}=\frac1{i\hbar}\big(H_\text{eff}\rho\SUB{\SS}-\rho\SUB{\SS} H^\dagger_\text{eff}\big)\delta t+\sum_{j=1}^{N-1}\big(J_j\rho\SUB{\SS} J_j^\dagger-r_j\rho\SUB{\SS}\big)\delta t\ .
\label{tee}
}
with the effective Hamiltonian
\EQ{
H_\text{eff}=H-\frac{i\hbar}2\sum_j\Big\{\sum_{\hat\mu}r_{j\hat\mu}F_{\hat\mu}^\dagger F_j+r_{Nj}F_j^\dagger-r_{jN}F_j+J_j^\dagger J_j-r_j\Big\}\ .
\label{hef}
}
We have assumed here that the Hilbert space of $\SS$ is finite dimensional, but we expect that it is possible to extend the argument to the infinite dimensional case.

The variation \eqref{tee} allows us to extract the dynamics of the conditioned state by writing the variation of $\rho\SUB{\SS}=\ket{\psi}\bra{\psi}$ as
\EQ{
\delta\rho\SUB{\SS}=\sum_{j=1}^N p_j\ket{\phi_j}\bra{\phi_j}-\rho\SUB{\SS}\ .
\label{tee2}
}
This yields
\begin{center}
\begin{tikzpicture}[scale=0.7]
\draw[fill=black] (1,0) circle (0.3cm);
\draw[very thick] (-1,0)  -- (1,0);
\draw[very thick] (1,0)  -- (3,1.5);
%\draw[thick] (0,0)  -- (3,0.5);
\draw[-] (1,0)  -- (3,0.3);
\node at (2.2,-0.3) {$\vdots$};
\draw[-] (1,0)  -- (3,-2);
\node at (-1.5,0) {$\ket{\psi}$};
\node[right] at (3.5,1.5) {$\ket{\phi_N}=\big(1+H_\text{eff}\,\delta t/i\hbar\big)\ket{\psi}$};
\node[right] at (3.5,-2) {$\ket{\phi_1}=J_1\ket{\psi}/\sqrt{r_1}$};
\node at (4.5,-0.8) {$\vdots$};
\node at (12.5,-0.8) {$\vdots$};
\node[right] at (3.5,0.3) {$\ket{\phi_{N-1}}=J_{N-1}\ket{\psi}/\sqrt{r_{N-1}}$};
\node[right] at (11.5,1.5) {$p_N=1-\sum_jr_j\delta t$};
\node[right] at (11.5,-2) {$p_1=r_1\delta t$};
\node[right] at (11.5,0.3) {$p_{N-1}=r_{N-1} \delta t$};
\end{tikzpicture}
\end{center}
Over the time interval $\delta t$, one of the probabilities is close to 1 and the probabilities of the other outcomes have a probability that is small, so there is a main ``trunk'' and a series of  ``branches''. This can be re-phrased as saying that during a small time interval $\delta t$ there is a probability $r_j\delta t$ for the instantaneous conditioned state $\ket{\psi}$ to branch out ---or jump ---into an orthogonal state $J_j\ket{\psi}$, for some operator $J_j$. The orthogonality of the $\ket{\phi_j}$, as required by Born's rule, follows from the orthogonality conditions \eqref{gut}.

\section{Absence of Schr\"odinger cat states}\label{a3}

In this appendix, we show why macroscopic superpositon states are rapidly localized to one of the components of the superposition with a probability given by the Born's rule. The discussion is mainly a review of S\"orgel and Hornberger \cite{SH}, but with some additional details. To this end, let us consider a state that is superposition of pointer states located at positions $x_i$:
\EQ{
\ket{\psi}\longrightarrow \sum_ic_i\ket{\psi_i}\ .
}
We will assume that the wave packets are narrow compared with the separation between the states. In this approximation, the state is completely specified by the positions of the centres $x_i$:
\EQ{
\langle x\rangle_\psi=\sum_iw_ix_i\ ,\qquad\Delta x_\psi^2=\sum_iw_i(x_i-\langle x\rangle_\psi)^2
}
where $w_i=|c_i|^2$. 

In order to understand the dynamics explicitly, let us consider the case of 2 wave packets. By substituting the superposition into the non-linear, non-Hermitian Schr\"odinger equation with Hamiltonian $H_\text{eff}$ \eqref{hef}, gives equations for the weights
\EQ{
\frac{dw_i}{dt}=2\Lambda L^2w_i(1-w_i)(2w_i-1)\ ,
\label{pyp}
}
where $L=|x_1-x_2|$. We can solve these equations explicitly; for $w_i=w_i(t)$ and $w_i'=w_i(t')$
\EQ{
w_i'=\frac12\Big(1+\text{sign}(2w_i-1)\sqrt{\frac y{4+y}}\Big)\ ,\qquad y=\frac{(1-2w_i)^2}{w_i(1-w_i)}\exp[2L^2\Lambda(t'-t)]\ .
}
It is clear from this that as $t\to\infty$ the largest of the $w_i$ goes to 1 while the smaller goes to 0. This would represent a violation of Born's rule were it not for the jumps which we have yet to consider.

\pgfdeclareimage[interpolate=true,width=6cm]{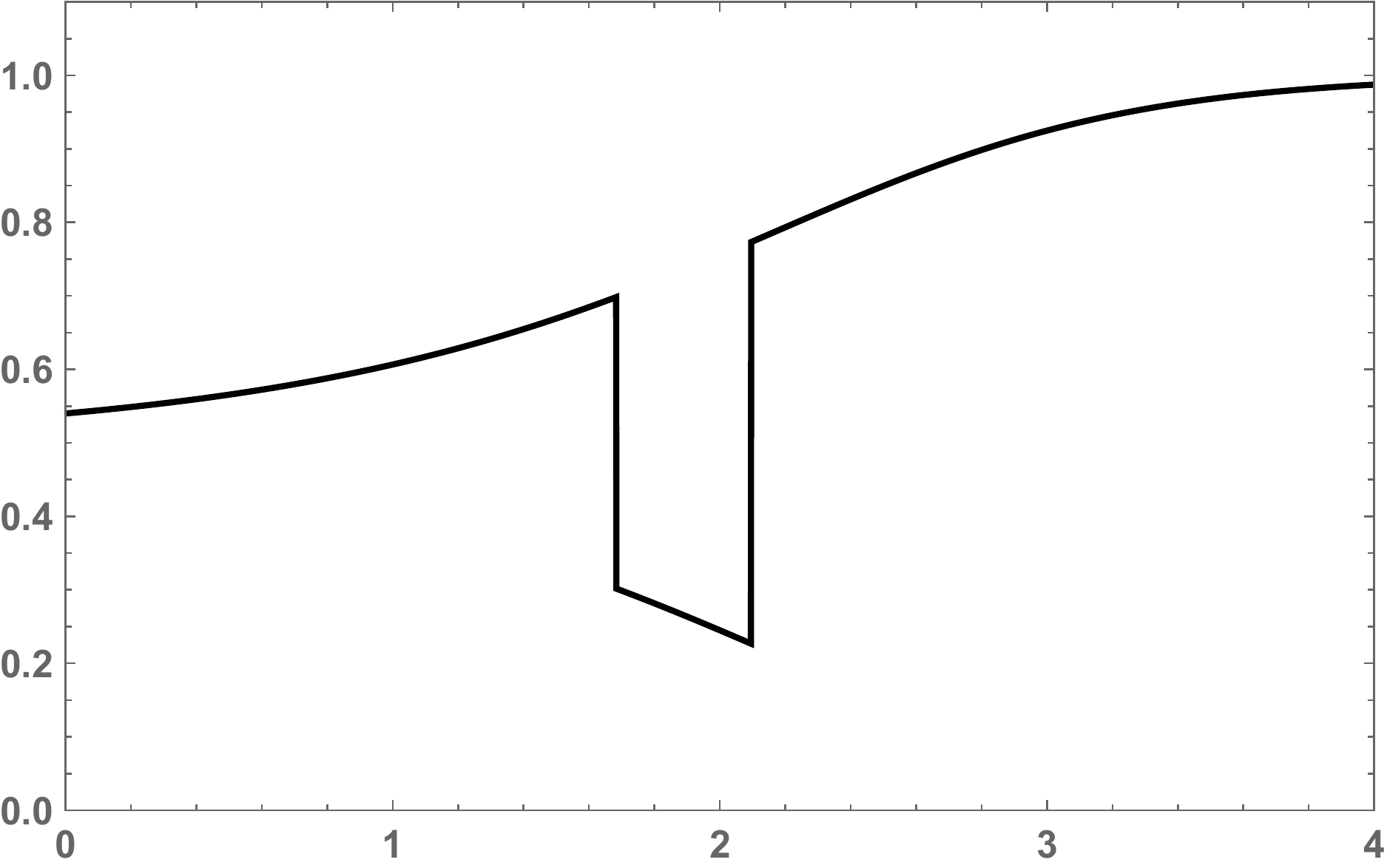}{f1}
\pgfdeclareimage[interpolate=true,width=6cm]{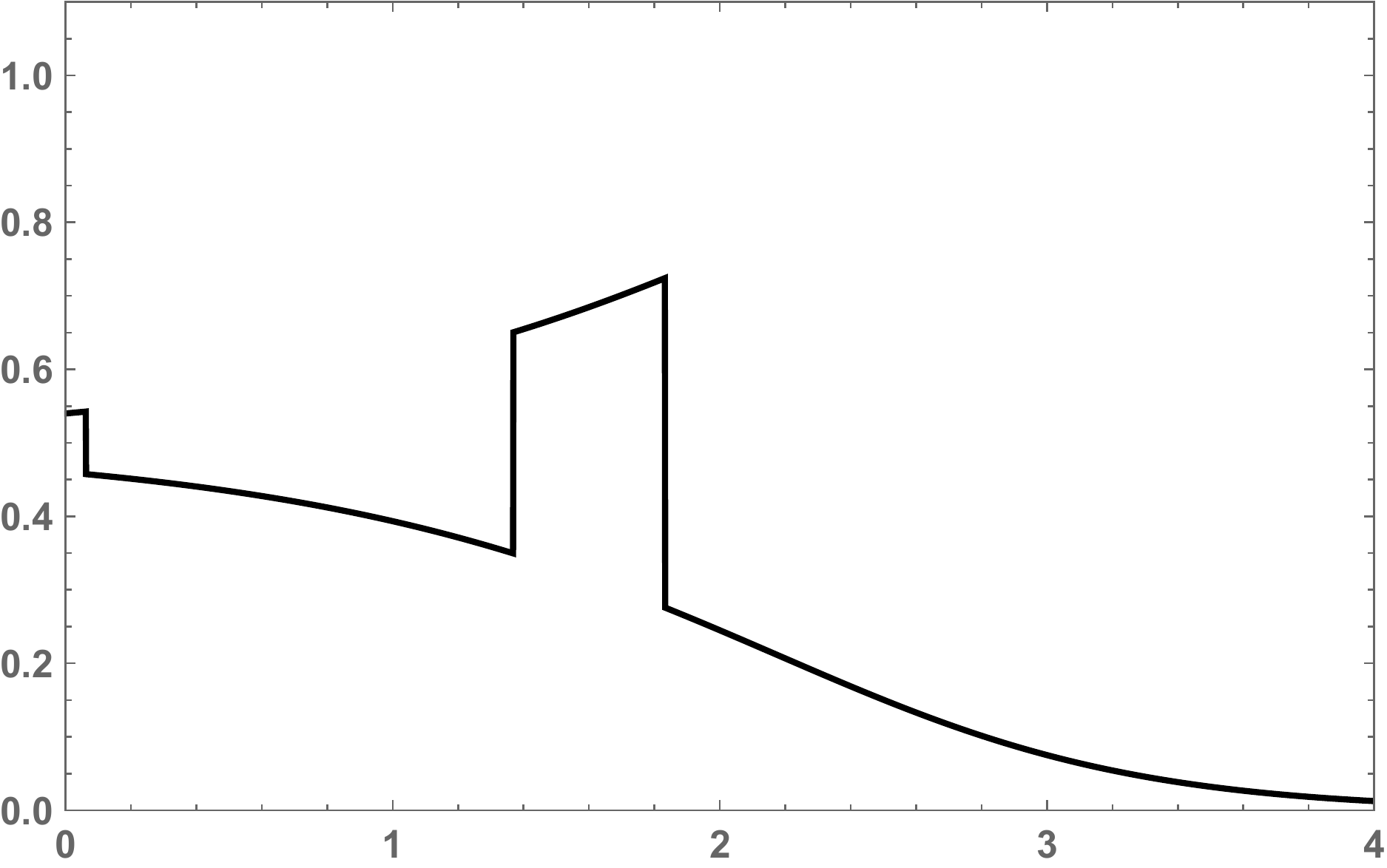}{f2}
\pgfdeclareimage[interpolate=true,width=6cm]{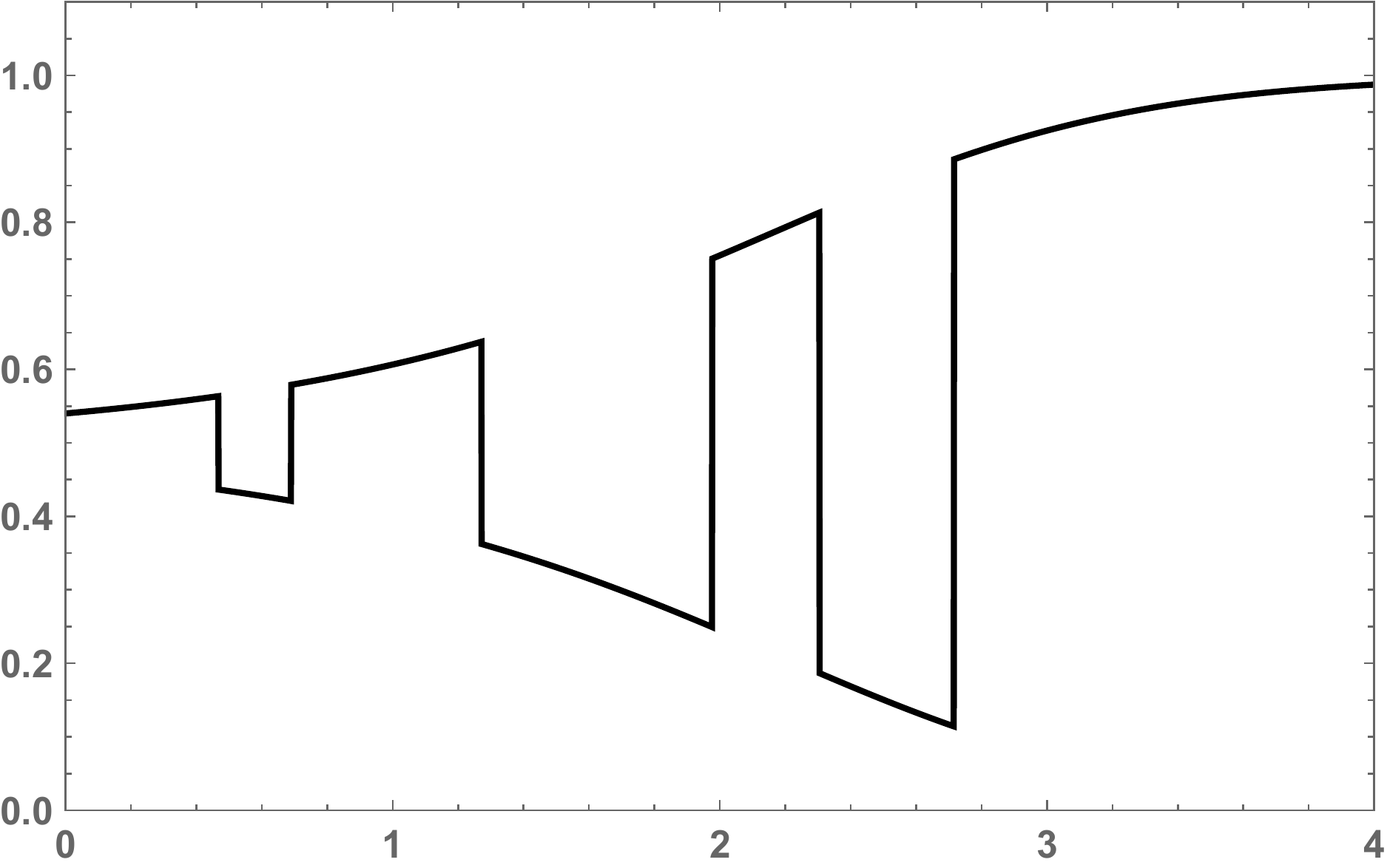}{f3}
\pgfdeclareimage[interpolate=true,width=6cm]{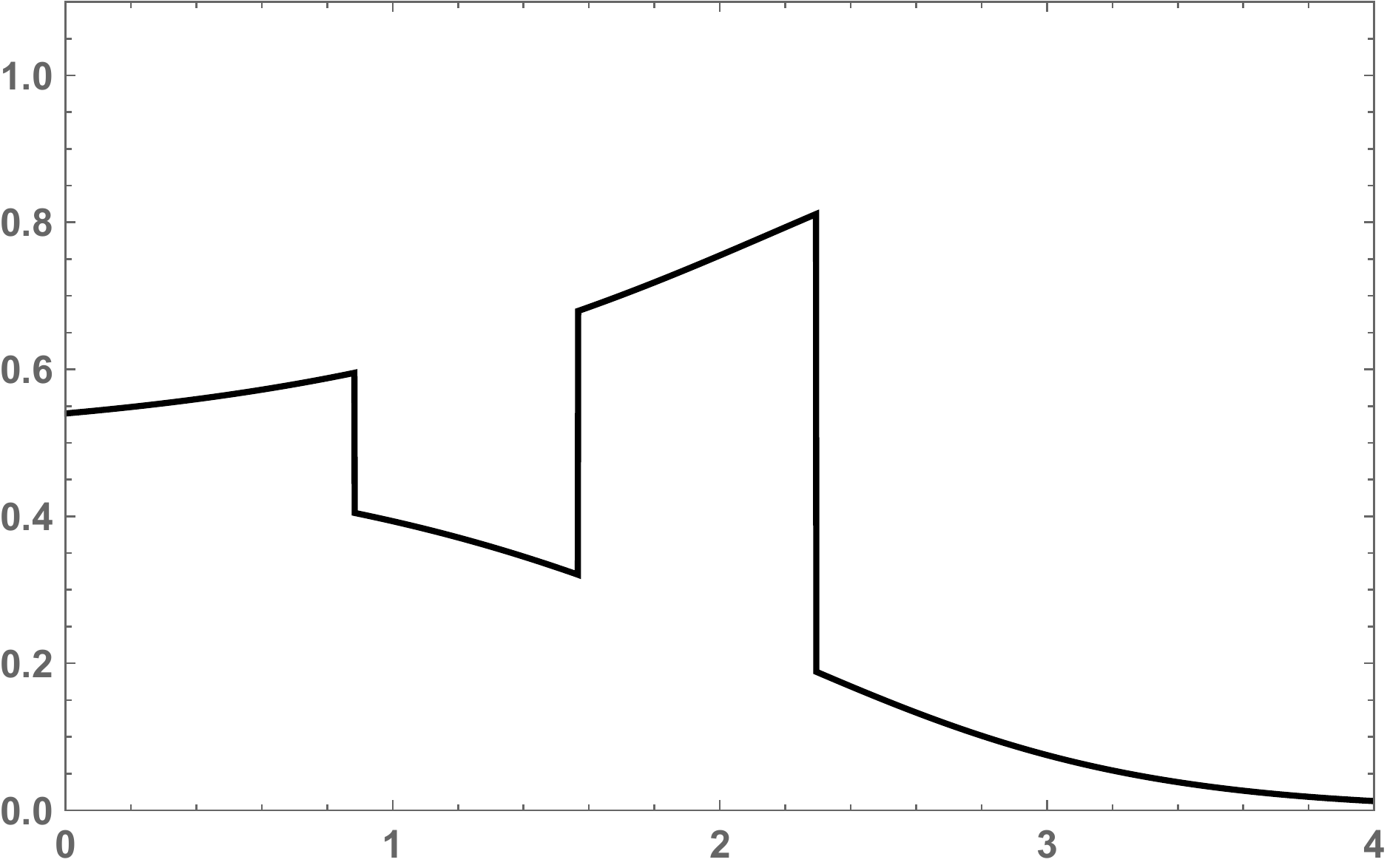}{f4}

\begin{figure}
\begin{center}
\begin{tikzpicture}[scale=1]
\pgftext[at=\pgfpoint{0cm}{0cm},left,base]{\pgfuseimage{f1}} 
\pgftext[at=\pgfpoint{7cm}{0cm},left,base]{\pgfuseimage{f2}} 
\pgftext[at=\pgfpoint{0cm}{-4.5cm},left,base]{\pgfuseimage{f3}} 
\pgftext[at=\pgfpoint{7cm}{-4.5cm},left,base]{\pgfuseimage{f4}} 
\end{tikzpicture}
\end{center}
\vspace{-0.5cm}
\caption{\footnotesize Four simulations of the trajectory of one of the weights which starts off $w_1(0)>1/2$. What is clear is that the rate of approach to localization is independent of the number of jumps. The latter increases as the initial condition $w_1(0)\to 1/2$.}
\label{fig2}
\end{figure}
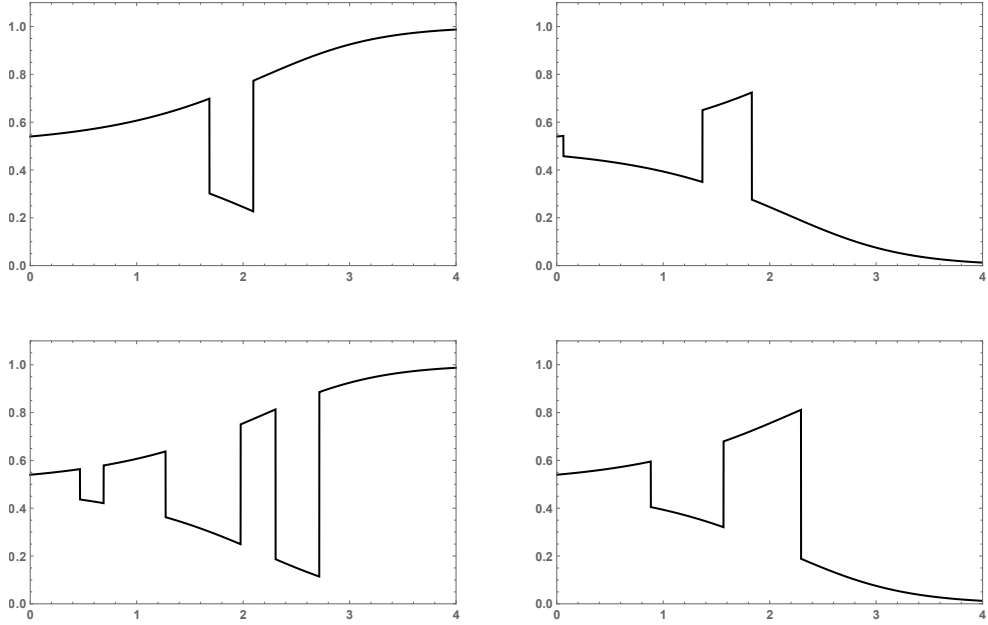

The rate of the jumps is
\EQ{
r=2\Lambda\Delta x_\psi^2=2\Lambda L^2w_1w_2\ .
}
In the collective coordinate approximation, the jumps correspond to a transformation of the weights $w_i$:
\EQ{
w_i\longrightarrow \frac{w_i(x_i-\langle x\rangle_\psi)^2}{\sum_jw_j(x_j-\langle x\rangle_\psi)^2}\ ,
}
which for 2 wave packets is particularly simple:
\EQ{
w_1\to w_2\ ,\qquad w_2\to w_1\ ,
\label{swa}
}
so $w_1$ and $w_2$ swap over. Note from \eqref{pup} that $r\,dt=|dw_i/(2w_i-1)|$ which is an expression invariant under \eqref{swa} and this allows us to calculate the mean number of jumps,
\EQ{
{\cal N}=\int_0^\infty r\,dt=\frac12\log\frac1{|w_1(0)-w_2(0)|}\ .
}
Also note that for a jump, the change in $w_1$ is $w_2-w_1=1-2w_1$ and \eqref{pup} can be written $dw_i/dt=r(2w_i-1)$. This means that the change in $w_i$ from the evolution by $H_\text{eff}$ is precisely cancelled by the average change in $w_i$ caused by the jumps. In other words the ensemble average of the variation $dw_i$ vanishes. Given that for $t\to\infty$, only one of $w_1$ or $w_2$ is equal to 1 and the other to 0, means that the probability of the final conditioned state being $\ket{\psi_i}$ is precisely equal to $w_i(0)$, i.e.~Born's rule is satisfied.  One important point is that the jumps do not affect the time it takes to reach the localized state.

Finally, the localization occurs over a time scale $1/L^2\Lambda$ which is very rapid and means that superpositions are destroyed long before the scale $L$ becomes macroscopic. Four simulations of one of the weights are shown in figure \ref{fig2}.

\end{document}